\documentclass [11pt]{article}
\usepackage{graphicx}
\usepackage{amsmath,amssymb} 
\usepackage{latexsym}
\usepackage{mathrsfs}
\usepackage{color}
\definecolor{red}{rgb}{1,0,0}

\makeatletter
\def\section{\@startsection {section}{1}{\z@}{-3.5ex plus -1ex minus
 -.2ex}{2.3ex plus .2ex}{\large\bf}}
\def\subsection{\@startsection{subsection}{2}{\z@}{-3.25ex plus -1ex
minus -.2ex}{1.5ex plus .2ex}{\normalsize\bf}}
\makeatother
\makeatletter

\@addtoreset{equation}{section}

\makeatother

\def\pslash{\raisebox{1pt}{$\slash$} \hspace{-7pt} p}

\def\Dslash{\hspace{3pt}\raisebox{1pt}{$\slash$} \hspace{-9pt} D}

\def\bea{\begin{eqnarray}} \def\eea{\end{eqnarray}}
\def\be{\begin{equation}} \def\ee{\end{equation}} \def\nn{\nonumber}
\def\a{& \hspace{-7pt}}  \def\Z{{\bf Z}}
 \def\ov{\overline}

\begin{document}

\thispagestyle{empty}

\begin{center}

\hfill SISSA-85/2010/EP \\

\begin{center}

\vspace*{0.5cm}

{\Large\bf Simple and Realistic Composite Higgs Models
\\ [2mm] in Flat Extra Dimensions}

\end{center}

\vspace{1.4cm}

{\bf Giuliano Panico$^{a}$,
Mahmoud Safari$^{b}$ and Marco
Serone$^{b,c}$}\\

\vspace{1.2cm}

${}^a\!\!$
{\em Institute for Theoretical Physics, ETH Zurich, 8093 Zurich, Switzerland}

\vspace{.3cm}

${}^b\!\!$
{\em International School for Advanced Studies (SISSA) and Istituto Nazionale di Fisica Nucleare (INFN), Via Bonomea 265, I-34136 Trieste, Italy} 

\vspace{.3cm}

${}^c\!\!$
{\em Abdus Salam International Center for Theoretical Physics (ICTP), Strada Costiera 11, I-34151 Trieste, Italy}

\end{center}

\vspace{0.8cm}

\centerline{\bf Abstract}
\vspace{2 mm}
\begin{quote}

We construct new composite Higgs/gauge-Higgs unification (GHU) models in flat space that overcome 
all the difficulties found in the past in attempting to construct models of this sort.
The key ingredient is the introduction of large boundary kinetic terms for gauge
(and fermion) fields.
We focus our analysis on the electroweak symmetry breaking pattern and the electroweak
precision tests and show how both are compatible with each other.
Our models can be seen as effective TeV descriptions of analogue warped models.
We point out that, as far as electroweak TeV scale physics is concerned, one can rely on
simple and more flexible flat space models rather than considering their unavoidably more
complicated warped space counterparts.
The generic collider signatures of our models are essentially undistinguishable from
those expected from composite Higgs/warped GHU models, namely a light Higgs, colored fermion
resonances below the TeV scale and sizable deviations to the Higgs and top coupling.

\end{quote}

\vfill

\newpage

\section{Introduction}

Many alternative scenarios of new physics beyond the Standard Model (SM) have been proposed
to address the gauge hierarchy problem. Among these, an intriguing idea is the
possibility of identifying the Higgs field with the internal component of a gauge field in extra dimensions \cite{early}, resulting in the
so called gauge-Higgs unification (GHU) models. In warped space  \cite{Randall:1999ee}, as suggested by the AdS/CFT duality, GHU models 
can be seen as a (relatively) weakly coupled 5D dual of 4D strongly
coupled conformal field theories~\cite{ArkaniHamed:2000ds}, where the Higgs field emerges
as a composite pseudo-Goldstone bound state of the strong sector~\cite{Contino:2003ve,Agashe:2004rs}.\footnote{
Contrary to the original AdS/CFT duality \cite{Ads-cft},
where both sides of the duality are well-defined, the 4D dual theories
of the GHU models in warped spaces are unknown. More precisely, what is so far
lacking is an UV description of the 4D theories in terms of fundamental states such as
quarks, gluons or strings, while we know, through the 5D construction, the low-energy
``chiral" Lagrangian associated to them.} From a wider perspective, GHU models in warped space are
concrete and successful realizations of the old idea \cite{Kaplan} of composite Higgs models.
The simplest composite Higgs models consist of two sectors:
an ``elementary'' sector, which includes the gauge and fermion fields of the Standard Model (SM), and
a ``composite'' sector, which is strongly coupled and is invariant under a
suitable global symmetry. The dynamics of the composite sector induces a spontaneous breaking of the global symmetry,
giving rise to a set of Goldstone bosons, that, for a judicious choice of the
symmetry group, can be identified with the Higgs field.
A small explicit breaking of the global symmetry is induced by gauging a part of it via the SM gauge bosons and by the weak mixing of the SM
fermions with the strong sector. The composite Higgs is thus a pseudo-Goldstone boson
and acquires a potential at the radiative level, which triggers electroweak symmetry breaking.

The symmetry structure of the composite Higgs scenario can be efficiently
used to perform some model-independent studies by using purely 4D low-energy effective
field theory considerations. This approach has been followed to find general
parametrizations of the non-linear sigma model describing the Higgs field
and its interactions~\cite{Giudice:2007fh} and of the interplay of the
SM fermions of the elementary sector with the composite sector~\cite{Contino:2006nn}.

Despite the importance of understanding general qualitative properties of composite
Higgs models by means of 4D effective field theory methods, these approaches
can not furnish a complete description of the composite Higgs scenario.
In particular, they do not allow to study all the properties of the strongly coupled
sector and they do not allow to compute quantities which are related to a UV
completion of the effective theory.
A more quantitative description of the composite Higgs scenario
is so far only possible by 
 constructing explicit GHU models in extra dimensions.
 These allow to extract all the relevant low-energy observables, including
the ones which are usually not computable in the 4D theories, namely the Higgs potential
and the detailed mass spectrum of the theory, which is a crucial ingredient in determining the electroweak precision parameters. 
The symmetries of the extra dimensional set-up can be made explicit by using a holographic effective description~\cite{Luty:2003vm,Barbieri:2003pr,Contino:2004vy,Panico:2007qd}, 
in which the ``elementary" sector of the composite Higgs models is identified with the field components localized
at one end-point of the extra-dimension (hereafter denoted UV brane), taken to be a segment, and the ``composite"  sector is identified with the remaining field components in the bulk.
In this way, it is manifest that the Higgs field can be equivalently seen as a set of pseudo-Goldstone bosons
coming from a spontaneous breaking of the extra dimensional gauge invariance and
that the theory in the low-energy regime reproduces the symmetry structure of the SM.

Constructing a realistic composite Higgs/GHU model is not an easy task. 
The most constraining electroweak bounds one should consider are given by the $T$ and $S$ parameters \cite{Peskin:1991sw}
and by the deviation $\delta g_b$ with respect to the SM value of the coupling between the left-handed (LH) bottom quark and the Z vector boson.
Couplings $g_{bt,R}$ between the right-handed (RH) top and bottom quarks with the $W^\pm$ vector bosons should also be taken into account, given the rather stringent experimental bounds on them of ${\cal O}(10^{-3})$ \cite{delAguila:2008iz}.
 Potentially deadly tree-level corrections to $T$ and $\delta g_b$ can be controlled
by appropriate custodial symmetries \cite{Agashe:2003zs,Agashe:2006at},
but  some tension with the experimental bounds still remains due to sizable one-loop corrections to $T$ and $\delta g_b$ coming from the fermion sector of these theories \cite{Carena:2006bn}.
Studying composite Higgs/GHU models in warped space is also not technically easy.
As a matter of fact, although a few 5D GHU models have been constructed 
so far \cite{Agashe:2004rs,Contino:2006qr,Medina:2007hz}, only in one model~\cite{Panico:2008bx}
(a modified version of a model introduced in \cite{Contino:2006qr} to accommodate
a Dark Matter candidate) one-loop corrections to $S$, $T$ and $\delta g_b$
(and the Higgs potential explicitly determined) have been analyzed and the 
ElectroWeak Precision Tests (EWPT) successfully passed. It is then important to look for other potentially
interesting models and possibly find different phenomenological features of the
composite Higgs scenario.

In the present work we want to point out that, as far as we are interested in the
low-energy phenomenology of the composite Higgs models,
we do not really need to consider the technically challenging warped models.
Instead, we can rely on the much simpler flat space implementations of
the GHU idea. The resulting models may still be reinterpreted as calculable 5D descriptions
of 4D strongly coupled composite Higgs models. This is guaranteed by the holographic
interpretation, which shows that the low-energy symmetries
of the theory are independent of the specific form of the 5D metric. The Goldstone
nature of the Higgs fields, as well as the phenomenology of the fermionic and gauge
sectors, are similar on flat and warped spaces.
We can also identify, through the 5D description, the low-energy ``chiral"
Lagrangian associated to a would-be strongly coupled 4D dual theory.
The only relevant ingredient that warped space adds to this view is the near-con\-for\-ma\-li\-ty
of the 4D strong sector. This is an important feature for what concerns the high-energy
running of the parameters of the theory and the generation of a hierarchy between
the electroweak scale and some high-energy scale, such as the Planck mass. However,
as far as electroweak symmetry breaking dynamics and collider phenomenology is concerned, 
these high-energy properties are not essential and can be
reliably omitted from an effective description.

Unfortunately, the simplest constructions of GHU models in flat space
(see \cite{Serone:2009kf} for an overview and for earlier references) turned out to
be not fully satisfactory (see e.g. \cite{Panico:2005dh}).
One of the reasons for this failure was the lack of some custodial protection
mechanism for the electroweak precision
parameters. If custodial symmetries are introduced, the situation improves but
this is still not enough to build realistic theories, since one gets too low top and Higgs masses. Another key ingredient are the
so called boundary kinetic terms (BKT) \cite{Carena:2002me}.
When these are introduced and taken to be large, potentially realistic models can be constructed.\footnote{The potential interest of large
BKT were already appreciated in \cite{Scrucca:2003ra}, but applied to a model
with $SU(3)$ gauge group, where the absence of a custodial symmetry led to large
tree-level corrections to the $T$ parameter. The possibility of getting realistic
flat space models with $SO(5)$ gauge group and large BKT was recently pointed out
in \cite{Serone:2009kf}.}
More in detail, we construct in this paper three different models, all based on the minimal gauge group $SO(5)\times U(1)_X$, with bulk fermions
in the fundamental or adjoint representation of $SO(5)$.
In all the models, large BKT at the UV brane for the gauge fields are introduced.
In two models, large BKT at the UV brane for the bulk fermions are also assumed.
We denote them FBKT$_{10}$ and FBKT$_5$ models, where $5$ and $10$
denote the $SO(5)$ representations of the fermion bulk multiplets. In the third and
last model no fermion BKT are introduced. This model is actually not new, but rather
a flat space adaptation \cite{Serone:2009kf} of a model introduced in \cite{Contino:2006qr}.
We denote it with the same acronym used in \cite{Contino:2006qr}, MCHM$_5$.
All the models successfully pass the EWPT,
as can be seen in figs.\ref{fig:bkt10mhalpha}, \ref{fig:bkt5mhalpha} and \ref{fig:mchm5mhalpha}.
As far as naturalness is concerned, the  MCHM$_5$ model is the one with the best performances,
with a fine-tuning roughly estimated at the $10\%$ level. This is around a few $\%$
in the FBKT$_{10}$ and FBKT$_5$ models. The LH top and 
bottom doublet and the RH top quark show a sizable degree of compositeness in all models.
 Considering flat space leads to a great technical simplification in model building and to
very explicit and significantly simpler expressions for various quantities
compared to the warped space case.
Moreover, the number of free parameters can be reduced and the fermion multiplet structure
in 5D can be simplified.

The structure of the paper is as follows. In section 2 we present the general framework
underlying all our models and the procedure used to compute the EWPT.
In sections 3, 4 and 5  we introduce respectively the FBKT$_{10}$, FBKT$_5$ and MCHM$_5$
models and the corresponding results.
In section 6 we conclude. We report in appendix A some simple analytic formulas for the
one-loop fermion contributions to $T$, $S$ and $\delta g_b$ that might help the reader
to understand how the EWPT are successfully passed in our models.

\section{General Framework}

\label{sec:framework}

All the models we consider in this paper share some common properties that are summarized below.
The bulk gauge group is taken to be $G=SU(3)_c\times SO(5)\times U(1)_X$. As well-known, $SO(5)$ is the smallest group containing
an $SU(2)$ custodial symmetry and giving rise to only one Higgs doublet. The subgroup $U(1)_X$ is necessary to reproduce the correct weak-mixing angle.
We denote by $g_5$ and $g_{5X}$ the 5D gauge coupling constants of $SO(5)$ and $U(1)_X$, respectively.
The unbroken group at $y=L$ is $H=SU(3)_c\times SO(4)\times U(1)_X\simeq SU(3)_c\times SU(2)_L \times SU(2)_R \times U(1)_X$. The unbroken group at $y=0$ is $H^\prime=SU(3)_c\times SU(2)_L\times U(1)_Y=G_{SM}$, where the hypercharge $Y$ is $Y=X+T_{3R}$. 

We work in the following in the ``holographic" basis for the gauge fields, namely we define the SM gauge fields as those which have SM couplings (with no deviations) 
to the elementary fermions (i.e. completely localized at $y=0$ \cite{Barbieri:2003pr}).  
We  use holographic techniques to efficiently compute the Higgs potential and tree-level corrections to electroweak observables (see \cite{Serone:2009kf} for an introduction to the basic holographic techniques used in models with extra dimensions).
In the ``holographic" unitary gauge $A_y =0$ \cite{Panico:2007qd}, the Higgs field is encoded in the sigma-model field  (see Appendix C of \cite{Serone:2009kf} for our  $SO(5)$ conventions):
\be
\Sigma = \exp\Big[\sum_{\hat a=1}^{4} i\frac{ \sqrt{2} t^{\hat a} h_{\hat a} }{f_\pi}\Big]\,, \ \ \ f_\pi=\frac{\sqrt{2}}{g_5\sqrt{L}}\,.
\label{Sigmaso5}
\ee
Neglecting the color $SU(3)_c$ factor,  the boundary conditions (b.c.) for the (non-canonically normalized) gauge fields are as follows:
\bea
F_{\mu y,L}^a & = & F_{\mu y,R}^a = F_{\mu y,X}= 0\,, \ \ A^{\hat a}_\mu=0\,,\ \ \ a=1,2,3\,, \hat a \in {\cal G}/{\cal H}\,, \hspace{1.1cm}  y=L,  \\
F_{\mu y,L}^a & = & F_{\mu y,R}^3+F_{\mu y,X} =0, \ \ 
A^{\hat a}_\mu=A^{1,2}_{\mu,R}= 0, \ \  A_{\mu,R}^3=A_{\mu,X} =B_\mu\,, \ \ y=0. \nn 
\eea
We introduce localized gauge kinetic terms at $y=0$ only. 
The EW gauge Lagrangian is
 \be
{\cal L}_g = {\cal L}_{5g}+{\cal L}_{4g,0}+{\cal L}_{4g,L},
\ee
with
\bea
{\cal L}_{5g} & = & \int_0^L \!\! dy\bigg\{ \frac{1}{2g_5^2}  {\rm Tr} \Big[-\frac 12 F_{\mu\nu}^2 + (\partial_y A_\mu)^2   \Big]+ \frac{1}{2g_{5X}^2}   \Big[-\frac 12 F_{\mu\nu,X}^2 + (\partial_y A_{\mu,X})^2  \Big]\bigg\} .\nn \\
{\cal L}_{4g,0} & = & -\frac{\theta L}{4g_5^2}  \sum_{a=1}^3 (W_{\mu\nu}^a)^2 - \frac{\theta^\prime L}{4g_{5X}^2}  B_{\mu\nu}^2\,, \ \ \ \ \ 
{\cal L}_{4g,L} = 0\,. \label{Lgauge}
\eea
In eq.(\ref{Lgauge}), $W_{\mu\nu}$ and $B_{\mu\nu}$ are the field strengths of the $SU(2)_L$ and $U(1)_Y$ gauge bosons, respectively, $\theta$ and $\theta^\prime$ are dimensionless parameters and the $SO(5)$ generators are normalized as ${\rm Tr}\, t_a t_b=\delta_{ab}$ in the fundamental representation.
We do not report the holographic Lagrangian for the SM gauge fields $W_\mu^a$ and $B_\mu$, that can be found  in \cite{Serone:2009kf}.
The SM gauge couplings constants $g$, $g^\prime$, and the Higgs VEV $v$ are related as follows to the 5D parameters:
\be
\frac{1}{g^2} \simeq \frac{L \Big(1+\theta\Big)}{g_5^2}\,, \  \  \
\frac{1}{g^{\prime 2}} \simeq \frac{L(1+\theta^\prime)}{g_{5X}^2} + \frac{L}{g_5^2}, \ \  \
v^2 = \frac{2 s_\alpha^2}{g_5^2 L}=f_\pi^2 s_\alpha^2,
\label{defgso5}
\ee
 valid for $\alpha \lesssim 1/3$, the region of interest. In eq.(\ref{defgso5}), 
 \be
 s_\alpha\equiv \sin(\alpha),\ \ \ \  \alpha=\frac{\langle h\rangle}{f_h} \,, \ \ \ \ \ \ 
 h=\sqrt{\sum_{\hat a=1}^4 h_{\hat a}^2} \,, \ \ \ \ \
 \langle h\rangle\simeq 246 \, {\rm  GeV}.
 \ee
In the holographic basis, the custodial $SU(2)_D$ symmetry, unbroken at $y=L$,
is completely manifest, resulting in a vanishing $T$ parameter at tree-level  \cite{Agashe:2003zs}. 
The $S$ parameter is not vanishing and given by 
\be
 S_{tree} \simeq \frac{4 s^2_W}{3\alpha_{em}}\frac{s_\alpha^2}{1+\theta} \,, \ \ \ \ \ 
\label{Streeso5}
\ee
where $\alpha_{em}$ is the electromagnetic constant at the $M_Z$ scale, $\alpha_{em}\simeq 1/129$, and
$s_W\equiv \sin\theta_W$, with $\theta_W$ the weak-mixing angle.
For $s_\alpha \lesssim 1/3$, the mass of the $W$ is given by
\be
M_W \simeq \frac{s_\alpha}{\sqrt{2} L \sqrt{\theta+1}}\,.
\label{mwSO5}
\ee
In the same limit, the mass $M_g$ of the lightest non-SM vector mesons is  
\be
M_g\simeq \frac{\pi}{2L}\,.
\label{MgaugeKK}
\ee 
The gauge contribution to the Higgs potential, for $\theta\sim \theta^\prime \gg 1$ and $s_\alpha \ll 1$, is well approximated by
\be
V_g \simeq \frac{3}{2} \int \!\!\! \frac{d^4p}{(2\pi)^4} \bigg[ 2\log\bigg(1+s_{\alpha}^2\frac{ \Pi_g^--\Pi_g^+}{2(\Pi_g^++\theta L p^2)}\bigg)+\log\bigg(1+ s_{\alpha}^2\frac{\sec^2 \theta_W ( \Pi_g^--\Pi_g^+)}{2 (\Pi_g^++\theta L p^2)}\bigg)\bigg]\,,
\label{Potso5gauge}
\ee
where 
\be
\Pi_g^+(p)=p \tan(pL), \ \ \ \  \Pi_g^-(p)= -p \cot(p L)\,.
\label{PiGauge}
\ee

Let us now turn to the model-dependent fermion sector of the Lagrangian. We only consider bulk fermions in the ${\bf 5}$ or ${\bf 10}$ representation of $SO(5)$. The fermion Lagrangian that encompasses all models has the following form:
 \be
{\cal L}_f = {\cal L}_{5f}+{\cal L}_{4f,0}+{\cal L}_{4f,L},
\ee
with
\bea
{\cal L}_{5f} & = & \int_0^L \!\! dy  \bigg[ \sum_{i=1}^{n_5} \bar\xi_i (i\Dslash-m_i) \xi_i  + \sum_{\alpha=1}^{n_{10}} {\rm Tr}\, \bar\xi_\alpha (i\Dslash -m_\alpha)\xi_\alpha \bigg] \,, \label{FermGenBulk} \\
{\cal L}_{4f,0} & = & \sum_n Z_n \bar\psi_n i\Dslash \psi_n \,,  \label{FermGen0} \\
{\cal L}_{4f,L} &   =  &  \sum_n \tilde m_n \bar \psi_n \tilde \psi_n +h.c. \,. \label{FermGenL}
\label{FermGen}
\eea
In eq.(\ref{FermGenBulk}) $\xi_i$ and $\xi_\alpha$ denote the bulk fermions in the ${\bf 5}$ and ${\bf 10}$ of $SO(5)$, respectively.
In  eq.(\ref{FermGen0})  $\psi_n$ denote the $SU(2)_L\times U(1)_Y$  chiral fermion components of the bulk multiplets that are not vanishing at $y=0$,
and $Z_n$ are their corresponding boundary kinetic terms. In  eq.(\ref{FermGenL})  $\psi_n$ and $\tilde \psi_n$ denote the $SO(4)\times U(1)_X$  chiral fermion components of the bulk multiplets that are not vanishing at $y=L$ and can mix through the mass terms $\tilde m_n$. The fermions $\psi_n$ and $\tilde \psi_n$ have 
classical dimension two in mass, like a 5D fermion, so that the BKT $Z_n$  have dimension one and the IR mass terms $\tilde m_n$ are dimensionless.
 No localized fermions are introduced.

Once the symmetry between the two-end points $y=0$ and $y=L$ is broken by the BKT, the end-point (UV brane) where the BKT are non-vanishing effectively defines the ``elementary'' sector of the composite Higgs model and the resulting models resemble more closely the analogue ones in warped space. 
This is the main reason why we have not introduced similar BKT at $y=L$ for gauge and fermion fields. 
This choice is quantum mechanically stable. If not introduced at a given scale, BKT at $y=L$ will appear through running effects \cite{Georgi:2000ks},
but with small coefficients $\sim g^2/(16\pi^2)$.
Large BKT are also quantum mechanically stable, since in the limit in which the BKT becomes
infinite, the zero mode of the Kaluza-Klein (KK) tower of the associated field becomes purely elementary and decouples from the massive
composite KK modes.\footnote{This can be easily seen by noticing that the shape of the zero-mode
wave function is independent of the BKT and, in the limit of infinite BKT,
its normalization goes to zero.}
Including large BKT for the fermions is thus natural and does not affect the cut-off $\Lambda$ of the
model. On the contrary, the gauge BKT have an impact on $\Lambda$ and tend to
lower it (for a discussion see e.g. \cite{Serone:2009kf}).
The difference with the fermions comes
from the fact that the gauge BKT determine the 4D gauge coupling constant.
If we fix the 4D gauge coupling, in order to obtain large values for the $\theta$
parameters we need to increase the 5D gauge coupling, thus lowering $\Lambda$.

As mentioned in the introduction, the most stringent bounds on 5D models of this sort come from the $S$ and $T$ parameters and by the deviation $\delta g_b$ to the $Zb_L\bar b_L$ coupling. 
In all our models we exploit the $\Z_2$ LR symmetry that allows to keep the tree-level correction to $\delta g_b$ under control \cite{Agashe:2006at}. 
We compute the latter by using the holographic approach. The main contribution to $\delta g_b$ arises from higher order operators with Higgs insertions, which give a contribution of ${\cal O}(\alpha^2)$.  Higher-order derivative operators are suppressed by the fermion masses or Z boson masses and are respectively ${\cal O}(m_b L)^2$ or ${\cal O}(m_Z L)^2\sim {\cal O}(\alpha^2/\theta)$, where eq.(\ref{mwSO5}) has been used in the second relation. For large BKT, $\theta \gg 1$, all higher derivative operators can be neglected and we can reliably set the momentum of all external fields  to zero. In this limit, the computation greatly simplifies and compact analytic formulas can be derived.

Since the symmetries protecting the $T$ parameter
and $Zb_L\bar b_L$ at tree-level are not exact, one-loop effects to such observables are expected to be important and must be included \cite{Carena:2006bn}. 
Non-SM fermions, being significantly lighter than non-SM vector mesons, play the dominant role, so it is a good approximation to just compute the one-loop fermion (top) contribution to $T$ and $\delta g_b$.\footnote{In the holographic basis of the gauge fields, $Zb_L\bar b_L$ has anyhow only fermion contributions, since the mixing in the gauge sector (i.e. the S-parameter) is rotated away \cite{Agashe:2003zs}.}
Performing one-loop computation using holographic techniques is not easy, so we resort here to the more standard KK approach.
Along the lines of \cite{Carena:2006bn}, we compute the masses and the Yukawa couplings mixing the lightest KK states with the top quark and by standard techniques compute the one-loop correction to $T$ and $\delta g_b$ \cite{Bamert:1996px}. We actually also compute one-loop corrections to $S$,\footnote{These one-loop fermion contributions to $S$ and $T$ refer to the standard, rather than holographic, basis, but the two practically coincide,  because 
one-loop corrections from light SM fields are negligible.} since the one-loop suppression factor $g^2/(16\pi^2)$ is partially compensated by the mild hierarchy between the masses of the lightest non-SM gauge and fermion states. Indeed, the one-loop contribution to $S$ given by a fermion (see appendix) is roughly ${\cal O}(N_c y^2 v^2/(4\pi M_f^2))$, where $y$ is a Yukawa coupling, $M_f$ a vector-like fermion mass and $N_c=3$ is the QCD color factor. Using eqs.(\ref{mwSO5}) and (\ref{MgaugeKK}), we can write the ratio between the one-loop and the tree-level correction to $S$ as follows:
\be
\frac{S_{1-loop}}{S_{tree}} \sim \frac{N_c y^2}{16\pi^2}\frac{M^2_g}{M^2_f}\,.
\ee
Given that typically $M_g^2 \gtrsim 10 M_f^2$, we see that one-loop corrections to $S$ cannot totally be ignored, although they play a sub-dominant role with respect to $T$ and $\delta g_b$.\footnote{The uncalculable contribution to $S$ due to physics at the cut-off scale is ${\cal O}(v/\Lambda)^2$. For $\Lambda \sim 10/L$ (see \cite{Serone:2009kf}), this is two orders of magnitude smaller than $S_{tree}$, and thus safely negligible.} 
We report in appendix \ref{app:STbbloop} analytic formulas for the new physics fermion contribution to $S$, $T$ and $\delta  g_b$ in the simplified case in which only one vector-like fermion ($SU(2)_L$ singlet, doublet or triplet) is relevant. 

Possibly dangerous $Wb_R\bar t_R$ couplings are generated at tree-level only in the ${\rm FBKT}_{10}$ model. In contrast to $\delta g_b$, one-loop corrections to
$g_{bt,R}$ are expected to be negligible, being suppressed by the small bottom Yukawa coupling.

We test our models by performing a combined $\chi^2$ fit expressed in terms of the $\epsilon_i$ parameters \cite{Altarelli:1990zd}, following \cite{Agashe}.
We use the following theoretical values for the $\epsilon_i$ parameters\footnote{We thank A. Strumia for providing us 
with the updated numerical coefficients entering in the $\epsilon_i$ and of the correlation matrix $\rho$, computed for $M_t^{\rm} = 173.1\ {\rm GeV}$.}
\bea
\epsilon_1&  = &  \big(5.64-0.86\;  lh \big) \times 10^{-3}+ \alpha_{em} T_{NP}\,, \nn \\
\epsilon_2 & = & \big(-7.10+0.16 \;  lh \big)\times 10^{-3} \,, \nn \\
\epsilon_3&  = &  \big(5.25+0.54\;  lh \big)\times10^{-3} + \frac{\alpha_{em}}{4\sin^2\theta_W} S_{NP}\,, \nn \\
\epsilon_b & = & -6.47 \times 10^{-3} - 2\delta g_{b,NP} \,, \label{epsTheo}
\eea
where $T_{NP}$, $S_{NP}$ and $\delta g_{b,NP}$, defined in eq.(\ref{STgbDef}), encode the new physics contribution without the SM one and
$lh \equiv \log M_{H,eff}/M_Z$, with the effective Higgs mass $M_{H,eff}$ defined as \cite{Barbieri:2007bh}\footnote{Notice that in eq.(\ref{mHeff}) we have replaced $\Lambda$, as taken in  \cite{Barbieri:2007bh}, with $1/L$, because
in GHU models by locality the Higgs contribution to $\epsilon_i$ is finite and saturated at the compactification scale $1/L$, rather than at the
cut-off of the theory $\Lambda$.}
\be
M_{H,eff} = M_H \Big(\frac{1}{M_H L}\Big)^{\sin^2\alpha}\,.
\label{mHeff}
\ee
The experimental values of the $\epsilon_i$, as obtained by LEP1 and SLD data \cite{:2004qh}, are  
\be
\hspace{2.5em}
\begin{matrix}
\epsilon_1^{exp}&  = &  ( 5.03\pm 0.93)\times 10^{-3} \,,\\
\epsilon_2^{exp}&  = &  ( -7.73\pm 0.95)\times 10^{-3}\,, \\
\epsilon_3^{exp}&  = &  ( 5.44\pm 0.87)\times 10^{-3} \,, \\
\epsilon_b^{exp}&  = &  ( -6.36\pm 1.3)\times 10^{-3}\,.
\end{matrix}
\hspace{3.5em}
\rho = \left( \begin{matrix}
1 \,\, \a 0.72\, \a 0.87\, \,\a -0.29 \cr
 0.72\,\, \a 1\,\, \a 0.46\,\, \a -0.26 \cr
 0.87\,\, \a  0.46 \,\, \a 1 \,\,\a -0.18 \cr
 -0.29\, \, \a -0.26\, \,\a -0.18 \,\,\a 1 \cr
 \end{matrix}\;
\right)\;.
\label{epsExp}
\ee
Finally, the $\chi^2$ function is defined as
\be
\chi^2 = (\epsilon_i -\epsilon^{exp}_i) (\sigma^{-1})_{ij}  (\epsilon_j -\epsilon^{exp}_j)\,, \ \ \  \sigma_{ij} = \sigma_i \rho_{ij} \sigma_j \,.
\label{chi2}
\ee

The  bound on $g_{bt,R}$ in the ${\rm FBKT}_{10}$ model is included by adding in quadratures to the $\chi^2$ (\ref{chi2}) the result coming from $b\rightarrow s \gamma$ decay \cite{delAguila:2008iz}:
\be 
g_{bt,R}  = (9 \pm 8) \times 10^{-4} \,.
\ee

Our results have been obtained by performing a random scan on the parameter space of
the models. The possibility of having simple analytic approximate formulas for the
top and bottom masses considerably helps in the scanning procedure, allowing us to
reduce the number of free parameters. In our analysis we take into account
in an approximate way the running of
the top mass by fixing its value at the energy scale $1/L$ in the range
$M_t(1/L) = (150 \pm 5)\, {\rm GeV}$.

In the next sections we will specify each model separately and present the results of our combined fit.

\section{Model I: ${\rm FBKT}_{10}$}

This is probably the simplest GHU model that can be built, with just one bulk multiplet $\xi$ in the adjoint representation ${\bf 10}$
of $SO(5)$. It is also the model with the least number of parameters we consider and probably the one with less parameters
so far in the literature. The ${\bf 10}$ decomposes as follows under $SO(4)$: ${\bf 10} = ({\bf 2},{\bf 2}) + ({\bf 1},{\bf 3})+({\bf 3},{\bf 1})$.
The boundary conditions of the LH components of  the multiplet $\xi$ are 
\begin{equation}
\xi_ L= \left( \begin{array}{c}\left\{
\begin{array}{c} x_{L} \, (+-)\\[0.05in] u_{L} \, (- -)\\[0.05in] d_{L} \, (--) 
\end{array}\right.\,\,\,\,\,  T_{L} \, (+-) \\[0.3in] 
\left[\, q^\prime_L \, (-+) \, , \, q_{L} \, (++) \, \right]
\end{array}
\right)_{\frac 23}\,,
\label{xi10bc}
\end{equation}
where the first and second entries in round brackets refer to the  $+$ ($-$) Neumann (Dirichlet) b.c. at the $y=0$ (UV) and $y=L$ (IR) branes, respectively. The RH components will have the opposite b.c., as usual. In eq.(\ref{xi10bc}), $q^\prime$ and $q$ are two
$SU(2)_L$ doublets, with $T_3^R(q^\prime)= 1/2$,  $T_3^R(q)= -1/2$, which form the $SO(4)$ bidoublet $({\bf 2},{\bf 2})$, $T_L$ is an $SU(2)_L$ triplet with $T_3^R(T)=0$ and the states in curly brackets are $SU(2)_L$ singlets forming a triplet of $SU(2)_R$.
The subscript $2/3$ denotes the $U(1)_X$ charge of the multiplet. We identify the
RH components of the top ($t_R$) and bottom ($b_R$) fields with the massless modes
of the $u_R$ and $d_R$ components respectively.

As explained in section \ref{sec:framework}, at $y=0$ we add the most general BKT for the non-vanishing field components there, namely
\begin{equation}
{\cal L}_{4f,0} = Z_q \bar q_L i\Dslash q_L + Z_t \bar u_R i\Dslash u_R
+ Z_b \bar d_R i\Dslash d_R + Z_x \bar x_L i\Dslash x_L +Z_T {\rm Tr}\, \bar T_L i\Dslash T_L
+ Z_{q^\prime} \bar q'_{R} i\Dslash q'_{R}. \label{FermGen0bkt10}
\end{equation}
No mass terms are allowed at $y=L$ and hence the IR localized Lagrangian is trivial:
\be
{\cal L}_{4f,L} = 0\,.
\ee

The holographic low-energy effective action, with the ``bulk''  physics integrated out, is up to ${\cal O}(s_\alpha^2)$ terms,
\be
{\cal L}_H = \bar q_L \frac{\pslash}{p} \Pi_0^q q_L
+ \sum_{a=t,b} \bar a_R  \frac{\pslash}{p} \Pi_0^a  a_R
 + \frac{s_{\alpha}}{h}\Big(
\Pi_M^t \bar q_L H^c t_R
+ \Pi_M^b \bar q_L H b_R+h.c.\Big) \,,
\label{LholoBKT10}
\ee
where 
\be
H =\frac{1}{\sqrt{2}} \left( \begin{matrix}
h_1-i h_2 \\
-h_3-i h_4 
\end{matrix}\right)\,, \ \ \ 
H^c \equiv i \sigma_2 H^\star =-\frac{1}{\sqrt{2}} \left( \begin{matrix}
h_3-i h_4 \\
h_1+i h_2 
\end{matrix}\right)\\.
\ee 
The explicit expression of the form factors appearing in eq.(\ref{LholoBKT10}) is the following:
\begin{eqnarray}
 \Pi_0^q &=& p Z_q + \Pi_+(c) \,,
\ \ \Pi_0^{t,b} = p Z_{t,b} - \frac{1}{\Pi_-(m)}\,,
\ \ \Pi_M^t = \frac{\Pi_M^b}{\sqrt{2}} = \frac{\Pi_-(m)-\Pi_+(m)}{\sqrt{2}\Pi_-(m)},
\hspace{1.5em}
\end{eqnarray}
in terms of the basic form factors
\begin{equation}
\Pi_+(m) = \frac{G_-(m)}{G_+(m)}\,, \qquad \Pi_-(m) = -\frac{G_+(-m)}{G_-(m)}\,,
\end{equation}
which, in turn, can be expressed 
in terms of the bulk to boundary fermion propagators
\begin{equation}
G_+(m) = \cos (\omega L) + \frac{m}{\omega} \sin (\omega L)\,, \qquad
G_-(m) = \frac{p}{\omega} \sin (\omega L)\,,
\label{bBprop}
\end{equation}
with $\omega \equiv \sqrt{p^2 - m^2}$.\footnote{Notice that the propagators $G_\pm$ in eq.(\ref{bBprop}) differ by a factor $\omega$ from those defined in \cite{Serone:2009kf}. In the form (\ref{bBprop}), the propagators are real also for imaginary values of $\omega$.}

Very simple formulas for the top and bottom masses can be obtained
by taking the zero momentum limit of the form factors appearing in the Lagrangian (\ref{LholoBKT10}). We have
\begin{equation}
\frac{M^2_{t}}{M^2_W} \simeq \frac{\theta+1}{2N_L N_{tR}}\,,\qquad
\frac{M^2_{b}}{M^2_W}  \simeq \frac{\theta+1}{N_L N_{bR}}\,,
\label{mtbBKT10}
\end{equation}
where
\bea
N_L & =  & \lim_{p\rightarrow 0} \frac{\Pi_0^q }{p L} = \frac{Z_q}{L}
+ \frac{1}{m L(\coth m L + 1)} , \nn \\
N_{tR,bR} & = &  \lim_{p\rightarrow 0} \frac{\Pi_0^{t,b}}{p L}  = \frac{Z_{t,b}}{L}
+ \frac{1}{m L(\coth m L - 1)}\,. \label{Norm0_bkt10}
\eea
Embedding a whole generation in a single bulk multiplet obviously implies that the upper
and lower non-canonically normalized Yukawa couplings in the holographic effective
Lagrangian (\ref{LholoBKT10}) are necessarily of the same order of magnitude,
in our case $|Y_b| = \sqrt{2} |Y_t|$.
Thus, the hierarchy between quark masses within a single generation does not arise from field localization in the extra dimension, but by demanding the $b_R$ to be more elementary than $t_R$, namely by taking the BKT $Z_b$ of $b_R$ much larger than the BKT $Z_t$ of $t_R$, so that $N_{bR}\gg N_{tR}$ and hence $M_{b}\ll M_{t}$. The spectrum of fermion resonances beyond the SM, before ElectroWeak Symmetry Breaking (EWSB), is given by KK towers of states in the ${\bf 2}_{7/6}$,  ${\bf 2}_{1/6}$,  ${\bf 1}_{5/3}$, ${\bf 1}_{2/3}$, ${\bf 1}_{-1/3}$ and ${\bf 3}_{2/3}$ of $SU(2)_L\times U(1)_Y$. 

\begin{figure}[t]
\begin{center}
\includegraphics[width=0.66\textwidth]{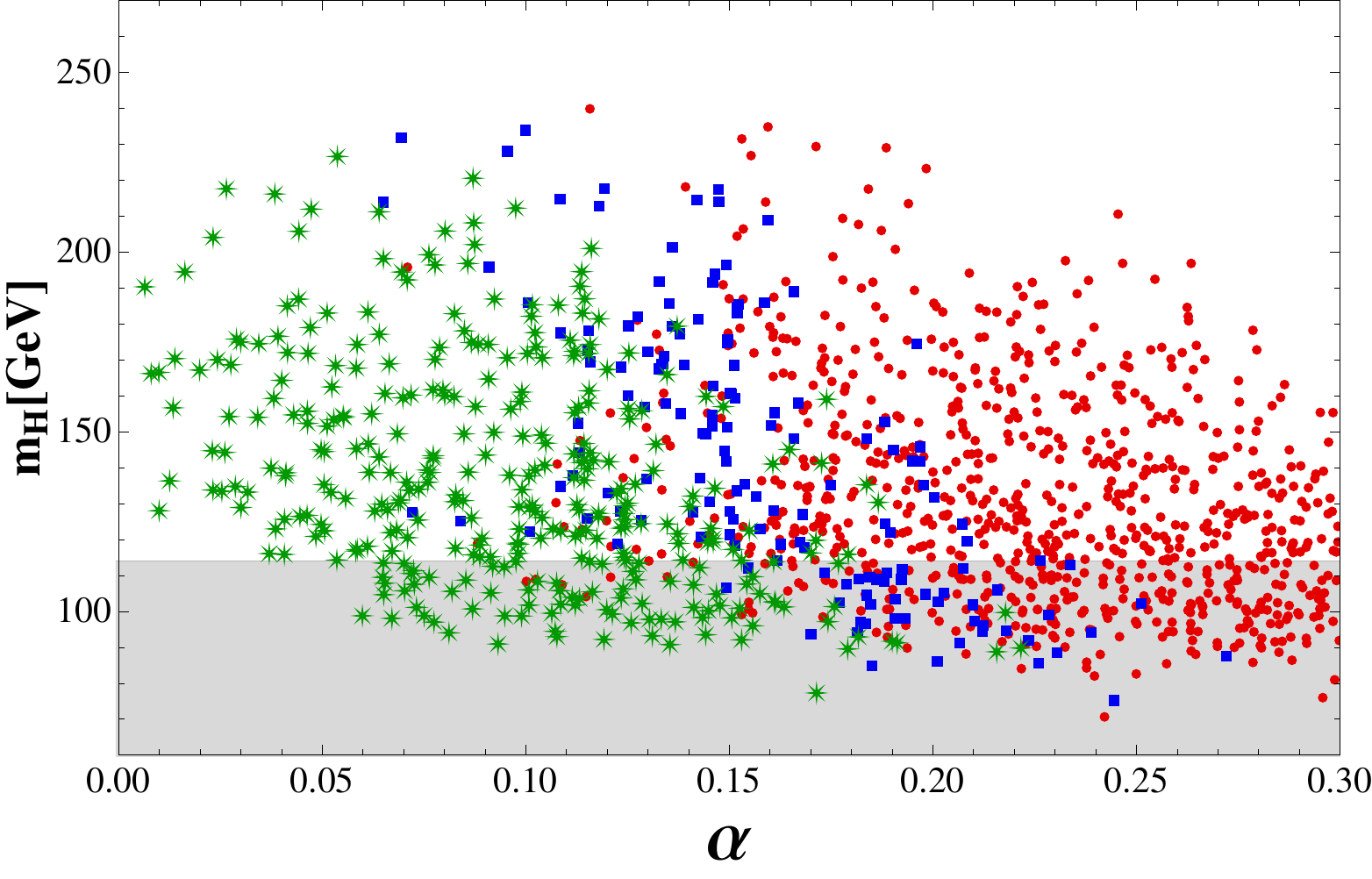}
\end{center}
\caption{\it Scatter plot of points obtained from a scan over the parameter space of the FBKT$_{10}$ model. Small red dots represent points which don't pass EWPT at $99\%$C.L.,
square blue dots represent points which pass EWPT at $99\%$C.L. but not at $90\%$ C.L.,
and star shape green dots represent points which pass EWPT at $90\%$C.L..
The region below the LEP bound ($m_H < 114\ {\rm GeV}$) is shaded. }
\label{fig:bkt10mhalpha}
\end{figure} 

The fermion contribution to the Higgs effective potential is the sum of three terms, coming from the states with $U(1)_Q$ charges $+5/3$, $+2/3$ and $-1/3$. The former contribution, $V_{ex}$, comes entirely from heavy states, while the latter two, $V_t$ and $V_b$, are related to the top and bottom KK tower of states. These contributions cannot be written in terms of the form factors appearing in eq.(\ref{LholoBKT10}), since higher order terms in $s_\alpha$ are missing  and, moreover, extra contributions arise from the bulk. The latter are absent only in the holographic basis where one chooses as holographic fields all the components of a multiplet with the same chirality \cite{Panico:2007qd}. This choice is manifestly not possible if we want
to keep the $q_L$, $t_R$ and $b_R$ components as holographic fields, given that they come from the same bulk field (\ref{xi10bc}).
The fermion contributions to the Higgs potential (mainly the top one) is quite lengthy.
For simplicity, we report in the following the explicit form of the Higgs potential only in the relevant region in parameter space where $Z_T,Z_q,Z_x\ll 1$ and $Z_b\gg 1$. 
Neglecting $Z_T$, $Z_q$ and $Z_x$, we get
 \bea
V_t & \simeq & -2 N_c \int \frac{d^4 p}{(2 \pi)^4} \ln\left(1 + s^2_\alpha 
\frac{(\Pi_--\Pi_+) \Big( 2p \Pi_+\Pi_- (Z_t-Z_{q^\prime})+\Pi_--\Pi_+\Big)}{4\Pi_+\Pi_- (p Z_{q^\prime}\Pi_+-1)(p Z_t \Pi_--1)}\right) \,, \nn \\
V_b & \simeq  & -2 N_c \int \frac{d^4 p}{(2 \pi)^4} \ln\left(1 + s^2_\alpha 
\frac{\Pi_+-\Pi_-}{2 p Z_b \Pi_+ \Pi_- }\right) \,, \nn \\
V_{ex} & \simeq  & -2 N_c \int \frac{d^4 p}{(2 \pi)^4} \ln\left(1 + s^2_\alpha 
\frac{\Pi_--\Pi_+}{\Pi_- (p Z_{q^\prime} \Pi_+ - 1)}\right) \,, \label{Vfermikonbkt10} 
\eea
where we have omitted the mass dependence of the form factors $\Pi_\pm$. The total Higgs potential is finally
\be
V_{tot} = V_g + V_t+V_b+V_{ex}\,,
\ee
with $V_g$ given in eq.(\ref{Potso5gauge}).

\begin{figure}[t]
\begin{center}
\includegraphics[width=0.66\textwidth]{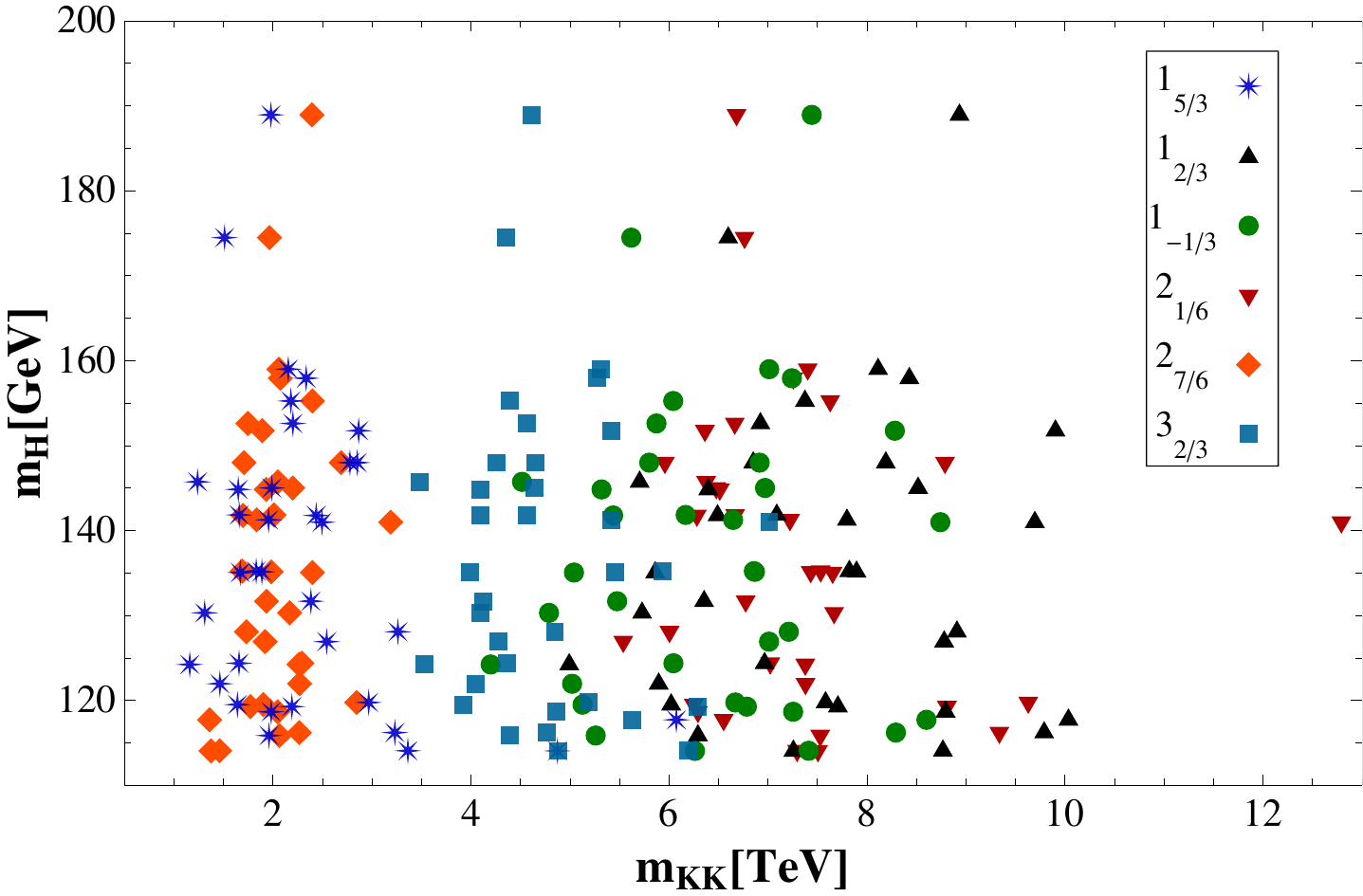}
\end{center}
\caption{\it Higgs mass $m_H$ versus the mass of the first KK resonances (before EWSB) for the points of the FBKT$_{10}$ model with $m_H>114\ {\rm GeV}$ and $\alpha\in[0.16,0.23]$ .}
\label{fig:bkt10mkk}
\end{figure} 

The tree-level contribution to $\delta g_{b}$ at leading order in an expansion in $\alpha$ is
\begin{equation}\label{Zbb_bkt10}
\delta g_{b} = \frac{e^{2 m L} m Z_T}{1-e^{2 m L}(1+2 m Z_q)}\frac{\alpha^2}{2}\,.
\end{equation}
The deviation (\ref{Zbb_bkt10}) crucially  depends on the BKT of the triplet, $Z_T$. When the latter vanishes, $\delta g_{b}=0$ (this is actually true to all orders in $\alpha$, at tree-level). This is a consequence of a $\Z_2$ custodial symmetry \cite{Agashe:2006at}.
More precisely, $\delta g_{b} \neq 0$ anytime, after EWSB,
$b_L$ sits in 5D fields where $T_3^R\neq T_3^L$ and the deviation is proportional to
$(T_3^R-T_3^L)$. In the case at hand, in absence of the BKT, there is a precise
cancellation between the contributions of the $T_3^R=-1, T_3^L=0$ and
of the $T_3^R=0, T_3^L=-1$ states. This compensation is explicitly broken by $Z_T$.
The mass-dependence of the result has also a simple physical interpretation.
Recall that in flat space, depending on the sign of the bulk mass term, KK states
with $(+-)$ or $(-+)$  b.c. become light exponentially, with an exponent governed by
the mass term $m$. For $m>0$, fermions with
$(+-)$ b.c. for the LH components become light, while for $m<0$ fermions with $(-+)$ b.c.
for the LH components become light.\footnote{This is completely analogous to the warped
space case, in which, due to a non-vanishing spin connection,
the relevant parameters are
$m/k\pm1/2$, where $k$ is the AdS$_5$ curvature scale.}
When $|mL|\gg 1$ and negative, the triplet tower is heavy and $\delta g_{b}$ is suppressed, while for positive $m$ the triplet tower becomes ultra-light and $\delta g_{b}$ is unsuppressed.

As we mentioned in section \ref{sec:framework}, in the FBKT$_{10}$ model the coupling $Wt_R b_R$ is generated at tree-level.
At the leading order in $\alpha$, we find
\be
g_{bt,R}= -\frac{\alpha^2}{2 \sqrt{2}} \frac{e^{2mL}-1}{2mL\sqrt{N_{tR}N_{bR}}}\,,
\ee
with $N_{tR}$ and $N_{bR}$ given in eq.(\ref{Norm0_bkt10}). When $Z_t=Z_b=0$, $g_{bt,R}$ is unsuppressed and it
equals
\be
g_{bt,R}= -\frac{\alpha^2}{2 \sqrt{2}} \,,
\ee
which is independent of $m$. This non-decoupling can heuristically be understood by noticing
that when $m<0$ the masses of the KK states mixing with $t_R$ and $b_R$ are large,
but  $t_R$ and $b_R$ are more composite (peaked toward the IR brane). On the contrary, when $m>0$, $t_R$
and $b_R$ are more elementary (UV peaked) but the KK states associated to the $q^\prime$ tower become
ultra-light. For any $m$, the two effects compensate each other, resulting in an
unsuppressed $g_{bt,R}$. When $Z_b$ and $Z_t$  are switched on, $g_{bt,R}$ is suppressed by
$N_{bR}$, which is required to be large to correctly reproduce the bottom mass.

\begin{figure}
\begin{center}
\includegraphics[width=0.6\textwidth]{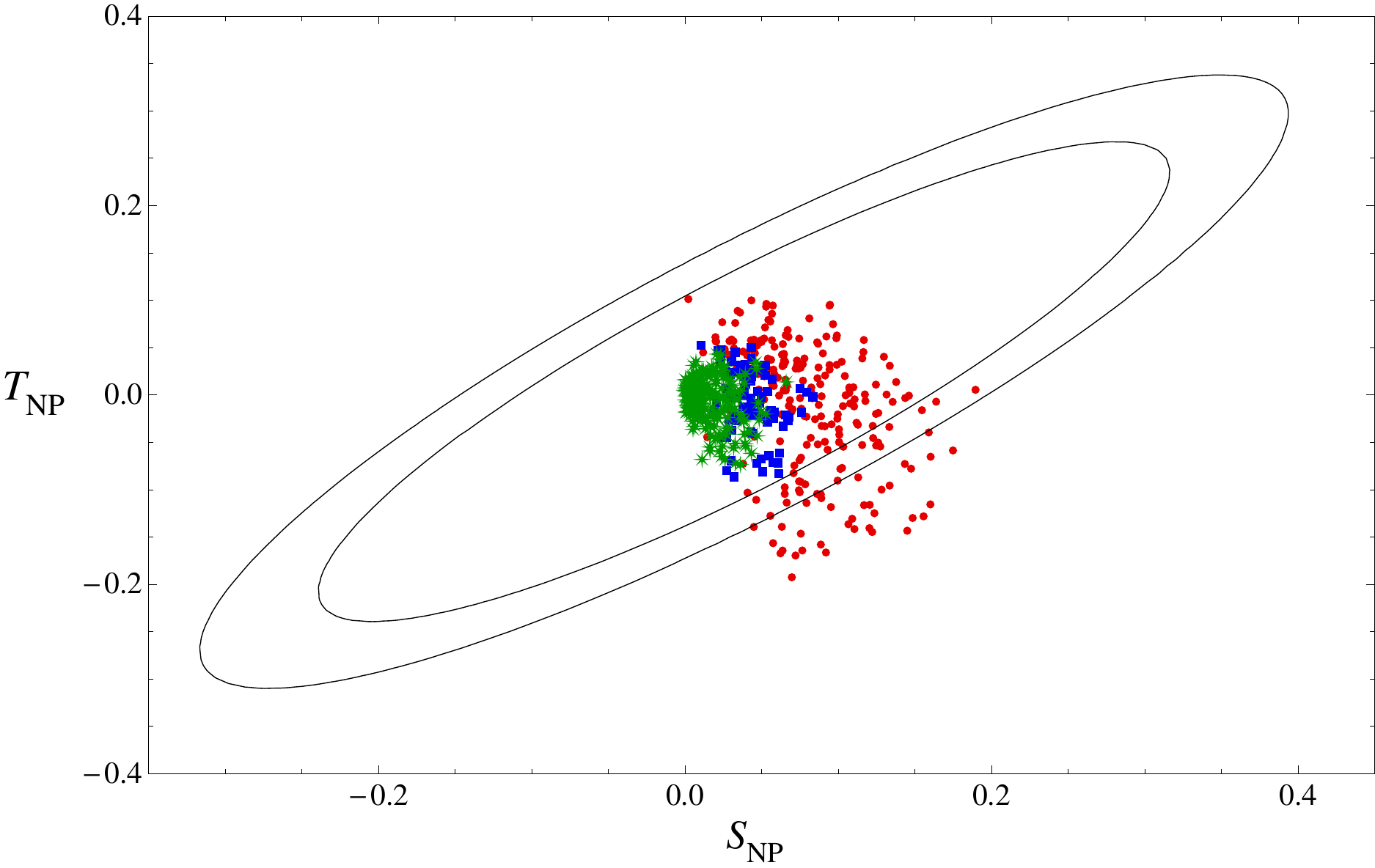}
\end{center}
\caption{\it Scatter plot of points in the FBKT$_{10}$ model with  $m_H > 114\ {\rm GeV}$ and  projected on the $T_{NP}$-$S_{NP}$  plane. We have set $M_{H,eff} = 120\ {\rm GeV}$. Small red dots represent points which don't pass EWPT at $99\%$ C.L., square blue dots represent  points which pass EWPT at $99\%$C.L. but not at $90\%$ C.L., and star shape green dots represent  points which pass EWPT at $90\%$ C.L.. The big and small ellipses correspond to $99\%$ and $90\%$ C.L. respectively.}
\label{fig:bkt10ST}
\end{figure}  

\subsection{Results}

The results of our numerical scan are summarized in figs.\ref{fig:bkt10mhalpha}, \ref{fig:bkt10mkk} and \ref{fig:bkt10ST}. 
The randomly chosen input parameters are $m$, $Z_{q}$, $Z_{q^\prime}$, $Z_x$, $Z_T$, $\theta$ and $\theta^\prime$.
The remaining two parameters $Z_t$ and $Z_b$ are fixed by the top and bottom mass formulas. For stability reasons, we take positive coefficients for all the BKT.
We have scanned the parameter space over the region $mL\in[-1.5,0.5]$, $Z_q/L\in[0,1.5]$, $Z_{q^\prime}/L\in[0,2]$, $Z_x/L\in[0,6]$, $Z_T/L\in[0,1.5]$, $\theta\in[20,30]$ and $\theta^\prime\in[15,25]$.

As can be seen in fig.\ref{fig:bkt10mhalpha}, the EWPT constrain $\alpha\lesssim 1/5$, with a light Higgs mass for the less-tuned points with $\alpha \simeq 0.15$. The Higgs mass increases only for more tuned configurations with $\alpha < 0.15$. The lightest exotic particles are fermion $SU(2)_L$ singlets with $Y=5/3$ and $SU(2)_L$ doublets with $Y=7/6$, see fig.\ref{fig:bkt10mkk}. After EWSB, these multiplets give rise to $5/3$ and $2/3$ charged fermions. 
Their mass is of order $1\div 2$ TeV, significantly lighter than the gauge KK modes ($\sim $ 5 TeV). The doublet  $q_L$ and the singlet $t_R$ have typically a sizable and comparable degree of compositeness, while $b_R$ is mostly elementary. When $mL\lesssim -1$, $q_L$ turns out to be even more composite than $t_R$.

\section{Model II: ${\rm FBKT}_{5}$}

The simplest model that can be built by using the fundamental representation of $SO(5)$
is constructed by embedding each generation of SM quarks in two bulk multiplets $\xi_t$ and
$\xi_b$. The ${\bf 5}$ decomposes as follows under
$SO(4)$: $\mathbf{5}=({\bf 2},{\bf 2})\oplus ({\bf 1},{\bf 1})$.
The boundary conditions on the fields,
\be
\begin{array}{c}
\begin{matrix}
\xi_{tL}= \left(
\begin{matrix}
(2,2)^{t}_L =
\begin{bmatrix}q^\prime_{1L}(-+) \\ q_{1L}(++) \end{bmatrix}
\\
(1,1)^{t}_L = u_{L}(--)
\end{matrix}\right)_{2/3},
& \ \ \ 
\xi_{bL}= \left(
\begin{matrix}
(2,2)^{b}_L =
\begin{bmatrix}q_{2L}(++) \\ q^\prime_{2L}(-+) \end{bmatrix}
\\
(1,1)^{b}_L = d_{L}(--)
\end{matrix}\right)_{-1/3},
\end{matrix} 
\end{array}
\label{fieldcomponents}
\ee
are fixed by the requirement of obtaining, out of each multiplet, the correct set
of massless components, namely a left-handed $SU(2)_L$ doublet and one $SU(2)_L$
right-handed singlet. For the third quark generation we can identify the $u_R$ and
$d_R$ zero-modes with the top and bottom RH singlets ($t_R$ and $b_R$).
On the other hand, the $q_{1L}$ and $q_{2L}$ zero-modes provide two copies of the LH
SM doublet and we need to eliminate a linear combination of the
two states from the massless spectrum. This can be easily done by modifying the
UV boundary conditions for the doublets and requiring a linear
combination of the two left-handed components to satisfy Dirichlet
conditions at the $y=0$ boundary (in our case we choose
$(q_{1L} + q_{2L})/\sqrt{2}$ as the SM doublet).\footnote{Equivalently one could
get rid of the unwanted massless doublet by
introducing a right-handed massless fermion doublet localized at the $y=0$ boundary,
which couples to the extra zero-mode with a large mass mixing.}

\begin{figure}[t]
\begin{center}
\includegraphics[width=0.66\textwidth]{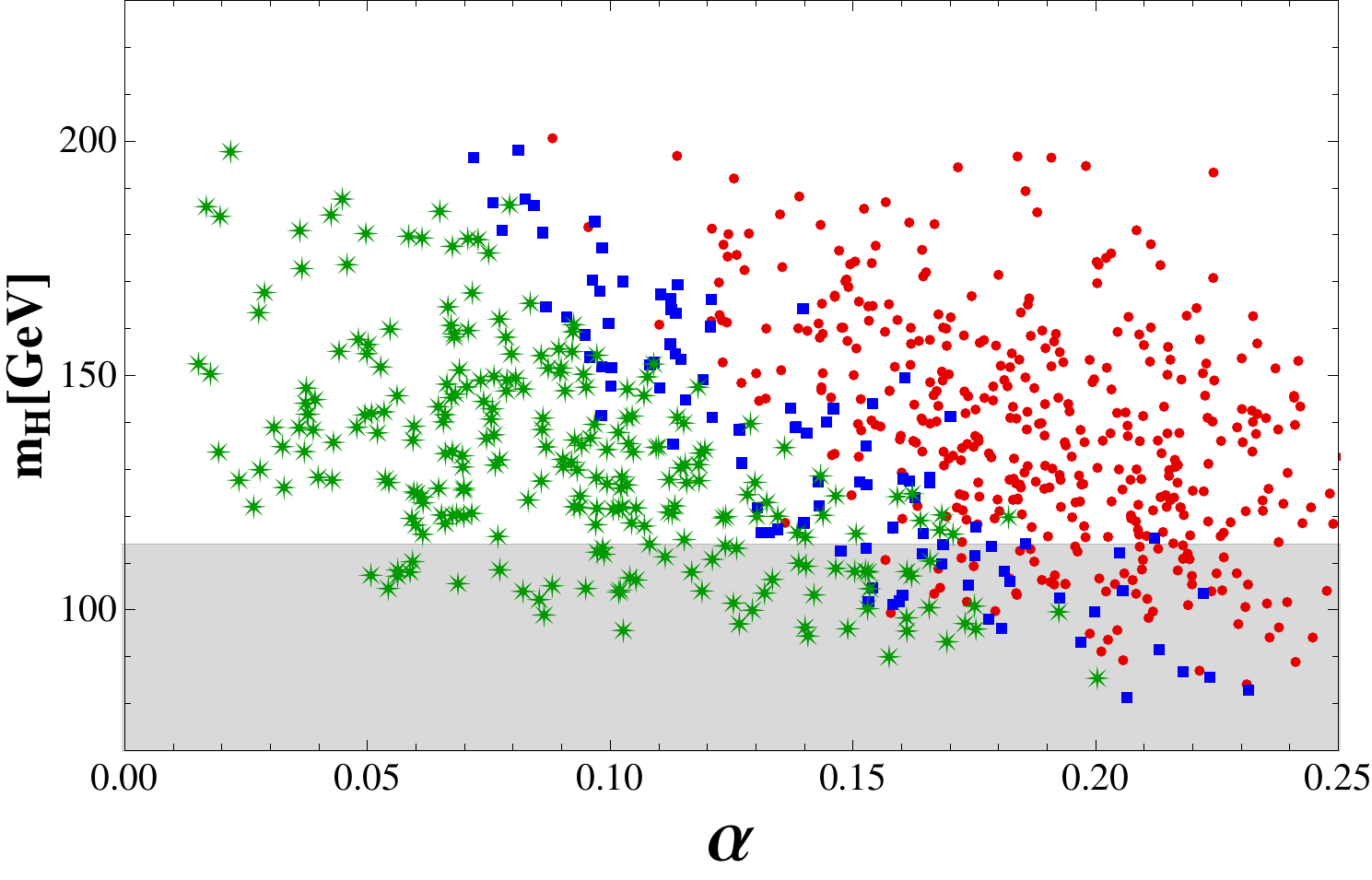}
\end{center}
\caption{\it Scatter plot of points obtained from a scan over the parameter space of the FBKT$_5$ model. Small red dots represent points which don't pass EWPT at $99\%$C.L.,
square blue dots represent points which pass EWPT at $99\%$C.L. but not at $90\%$C.L.,
and star shape green dots represent points which pass EWPT at $90\%$C.L..
The region below the LEP bound ($m_H < 114\ {\rm GeV}$) is shaded.}
\label{fig:bkt5mhalpha}
\end{figure} 

The most general BKT for the non-vanishing field components at $y=0$ are
\begin{equation}
{\cal L}_{4f,0} = Z_q \bar q_L i\Dslash q_L + Z_t \bar u_R i\Dslash u_R
+ Z_b \bar d_R i\Dslash d_R + Z_{R1} \bar q'_{1R} i\Dslash q'_{1R}
+ Z_{R2} \bar q'_{2R} i\Dslash q'_{2R}\,. \label{FermGen0bkt5}
\end{equation}
No mass terms are allowed at $y=L$ and hence the IR localized Lagrangian is trivial:
\be
{\cal L}_{4f,L} = 0\,.
\ee
The holographic low-energy effective action is, up to ${\cal O}(s_\alpha^2)$ terms, 
 of the form (\ref{LholoBKT10}), where 
\begin{eqnarray}
\Pi_0^q &=& p Z_q + \frac{1}{2}(\Pi_+(m_t) + \Pi_+(m_b))\,,\\
\Pi_0^{t,b} &=& p Z_{t,b} - \frac{1}{\Pi_-(m_{t,b})}\,,\\
\Pi^{t,b}_M &=& \frac{\Pi_-(m_{t,b}) - \Pi_+(m_{t,b})}{\sqrt{2} \Pi_-(m_{t,b})}\,.
\end{eqnarray}
Simple approximate formulas for the top and bottom masses are obtained from the Lagrangian (\ref{LholoBKT10}). One has
\begin{equation}
\frac{M^2_{t}}{M^2_W} \simeq \frac{\theta+1}{2N_L N_{tR}}\,,\qquad
\frac{M^2_{b}}{M^2_W}  \simeq \frac{\theta+1}{2N_L N_{bR}}\,,
\end{equation}
where 
\bea
N_L & =  & \lim_{p\rightarrow 0} \frac{\Pi_0^q }{p L} = \frac{Z_q}{L}
+ \frac{1 - e^{-2 L m_t}}{4 L m_t} + \frac{1 - e^{-2 L m_b}}{4 L m_b}\,, \nn \\
N_{tR,bR} & = &  \lim_{p\rightarrow 0} \frac{\Pi_0^{t,b}}{p L}  = \frac{Z_{t,b}}{L}
+ \frac{e^{2 L m_{t,b}} - 1}{2 L m_{t,b}}\,. \label{Norm0_bkt5}
\eea
The spectrum of fermion resonances beyond the SM, before EWSB, is given by KK towers of states in the ${\bf 2}_{7/6}$,  ${\bf 2}_{-5/6}$,  ${\bf 2}_{1/6}$,  ${\bf 1}_{2/3}$ and ${\bf 1}_{-1/3}$ of $SU(2)_L\times U(1)_Y$. 
\begin{figure}[t]
\begin{center}
\includegraphics[width=0.66\textwidth]{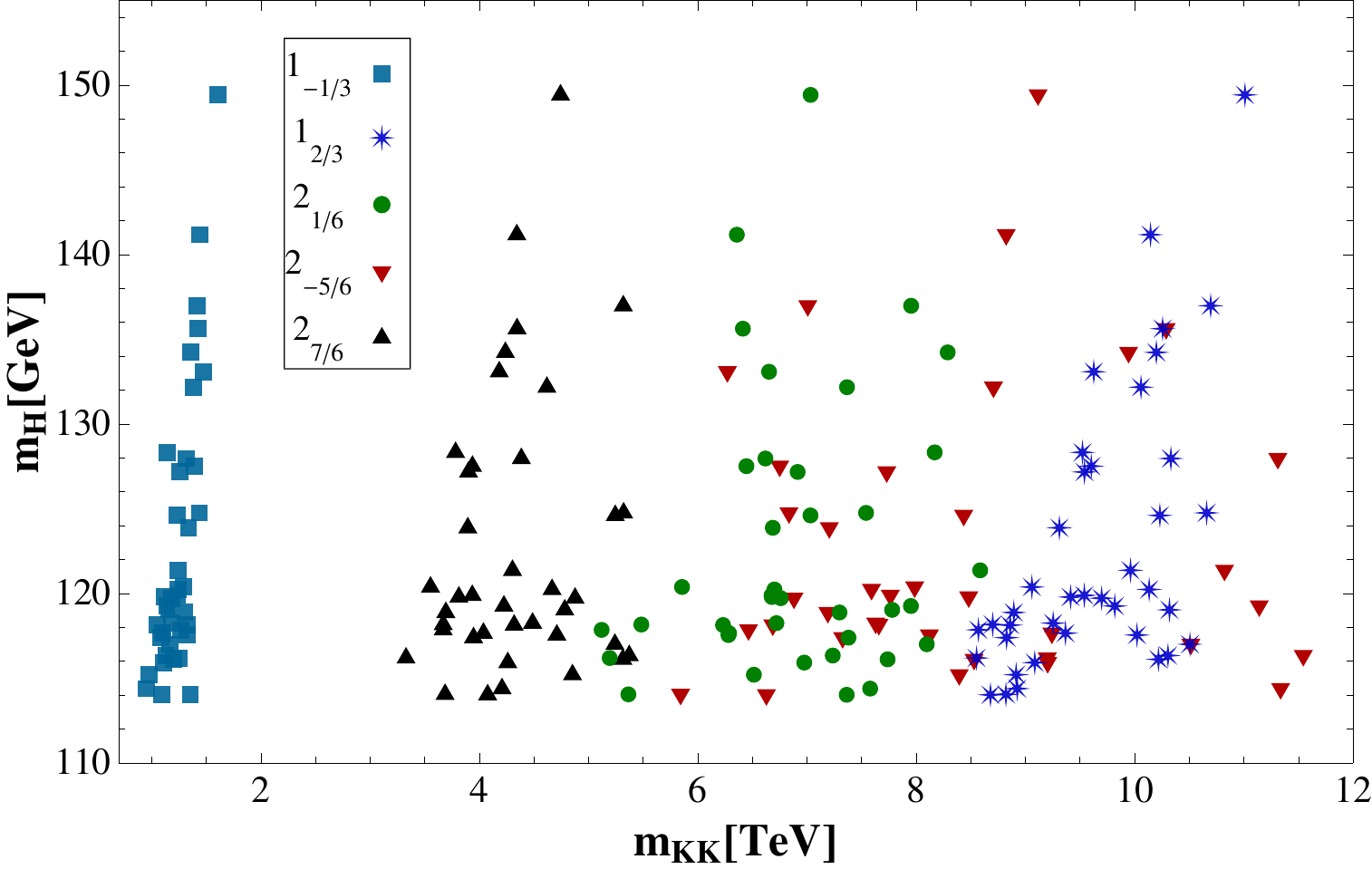}
\end{center}
\caption{\it  Higgs mass $m_H$ versus the mass of the first KK resonances (before EWSB) for the points of the FBKT$_{5}$ model with $m_H>114\ {\rm GeV}$ and $\alpha\in[0.16,0.22]$.}
\label{fig:bkt5mkk}
\end{figure}

The fermion contribution to the Higgs effective potential cannot  be written in terms of the form factors appearing in eq.(\ref{LholoBKT10}) 
for the same reasons explained for the FBKT$_{10}$ model above eq.(\ref{Vfermikonbkt10}). The fermion contribution to the Higgs potential  comes from the top and bottom
tower of states, $V_f = V_t + V_b$.
The explicit form of the top tower contribution to the potential is given by
\begin{eqnarray}
V_t &=& -2 N_c \int \frac{d^4 p}{(2 \pi)^4} \ln\left[1 + \sin^2 \alpha
\frac{\Pi_+(m_t)-\Pi_-(m_t)}{2(p Z_t \Pi_-(m_t) - 1)}
\left(p Z_t + p \frac{Z_{R1} - Z_t}{p Z_{R1} \Pi_+(m_t) - 1}\right.\right. \nn\\
&& \hspace{8em}\left.\left.- \frac{p Z_{t} \Pi_+(m_t) - 1}{2 p Z_q + \Pi_+(m_t) + \Pi_+(m_b)}\right)
\right]\,,
\end{eqnarray}
while the bottom tower contribution $V_b$ is obtained from $V_t$ by the
replacements $t \leftrightarrow b$ and $Z_{R1} \rightarrow Z_{R2}$.
The total Higgs potential is finally
\be
V_{tot} = V_g + V_t+V_b\,,
\ee
with $V_g$ given in eq.(\ref{Potso5gauge}).

\begin{figure}
\begin{center}
\includegraphics[width=0.6\textwidth]{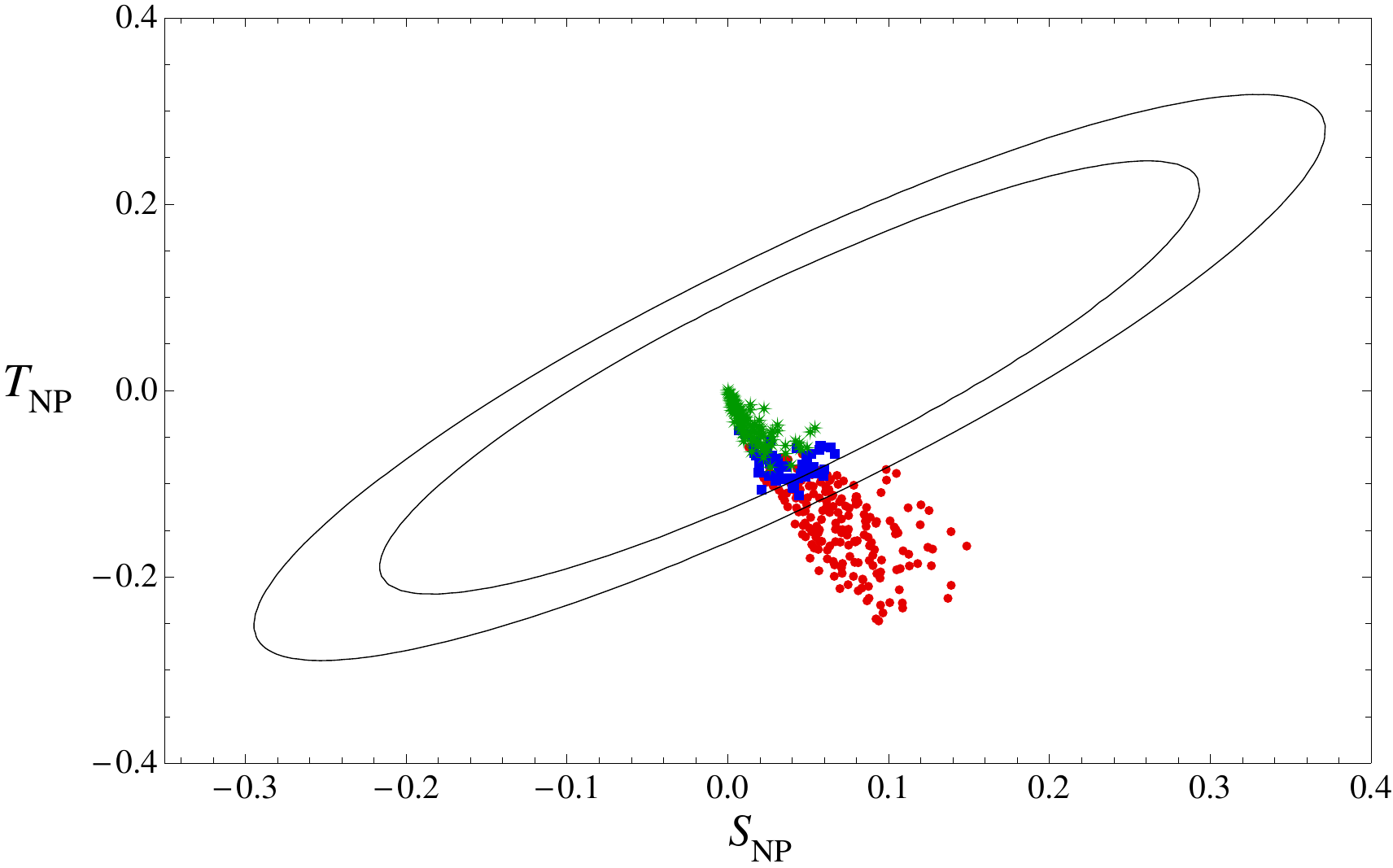}
\end{center}
\caption{\it Scatter plot of points in the FBKT$_5$ model with $m_H> 114\ {\rm GeV}$ and  projected on the  $T_{NP}$-$S_{NP}$ plane. We have set $M_{H,eff} = 120\ {\rm GeV}$. Small red dots represent points which don't pass EWPT at $99\%$ C.L., square blue dots represent points which pass EWPT at $99\%$C.L. but not at $90\%$ C.L., and star shape green dots represent  points which pass EWPT at $90\%$ C.L.. The big and small ellipses correspond to $99\%$ and $90\%$ C.L. respectively.}
\label{fig:bkt5ST}
\end{figure}

The tree-level contribution to $\delta g_{b}$ at leading order in an expansion in $\alpha$ is
\begin{equation}\label{Zbb_bkt5}
\delta g_{b} = \frac{1- e^{-2 m_b L}}{16 m_b L N_L} \alpha^2\,.
\end{equation}
The result (\ref{Zbb_bkt5}) has a simple physical interpretation. According to the analysis
of~\cite{Agashe:2006at}, multiplets in which the bottom lives only in components
with $T_3^R = T_3^L$ do not contribute to $\delta g_{b}$. This condition is
satisfied by the $\xi_t$ multiplet, hence the only corrections come from the $\xi_b$
field. A comparison with eq.~(\ref{Norm0_bkt5}) shows that, as expected,
the correction to $g_{b_L}$ is an ${\cal O}(\alpha^2)$ effect and is proportional
to the fraction of the $b_L$ wave function which lives in the $\xi_b$ multiplet,
which is encoded in the ratio on the right hand side of eq.~(\ref{Zbb_bkt5}).

\subsection{Results}

The results of our numerical scan are summarized in figs.\ref{fig:bkt5mhalpha}, \ref{fig:bkt5mkk} and \ref{fig:bkt5ST}.
The randomly chosen input parameters are $m_t$, $m_b$, $Z_{R1}$, $Z_{R2}$, $Z_q$, $\theta$ and $\theta^\prime$.
The remaining two parameters $Z_b$ and $Z_t$ are fixed by the top and bottom mass formulas. For stability reasons, we take positive coefficients for all the BKT
and $m_b L\gtrsim 1$ in order to suppress $\delta g_{b}$, as given by  eq.(\ref{Zbb_bkt5}). 
More precisely, we have taken $m_t L\in[0.1,1.3]$, $m_b L\in[2,2.5]$, $Z_{R1}/L\in[0.1,1.6]$, $Z_{R2}/L\in[0,1]$, $Z_q/L\in[0.5,2]$, $\theta\in[15,25]$, $\theta^\prime\in[15,25]$.
As can be seen in fig.\ref{fig:bkt5mhalpha}, the EWPT constrain $\alpha\simeq 1/5$, with a very light Higgs mass. The latter increases only for more tuned configurations with
$\alpha < 0.15$. Interestingly enough, the lightest exotic particle is always a fermion singlet with $Q=-1/3$, see fig.\ref{fig:bkt5mkk}. Its mass is of order 1 TeV, significantly lighter
than the gauge KK modes ($\sim $ 5 TeV) and the other fermion resonances, with masses starting from around 4 TeV.  The doublet  $q_L$ is generically semi-composite, the singlet $t_R$ is mostly composite and $b_R$ is mostly elementary.

\section{Model III: modified ${\rm MCHM}_{5}$}

\label{mchm5}

The last model we consider is the flat space version of one of the models considered in \cite{Contino:2006qr}
and denoted there MCHM$_5$. It was already noticed in \cite{Serone:2009kf} that this model can lead to realistic theories also when defined on a flat segment, provided large BKT for the gauge fields are included. Here we perform a systematic analysis of the electroweak bounds in this model, that was neither made in \cite{Serone:2009kf} nor in \cite{Contino:2006qr}. 
We describe very briefly the model, referring the reader to section 4.2 of \cite{Serone:2009kf}
or to the original warped space version \cite{Contino:2006qr} for further details. Fermion BKT can also be introduced in this model, of course, but given the
larger number of parameters present in this model with respect to the FBKT models, we have decided, for simplicity, to neglect them.

\begin{figure}[t]
\begin{center}
\includegraphics[width=0.66\textwidth]{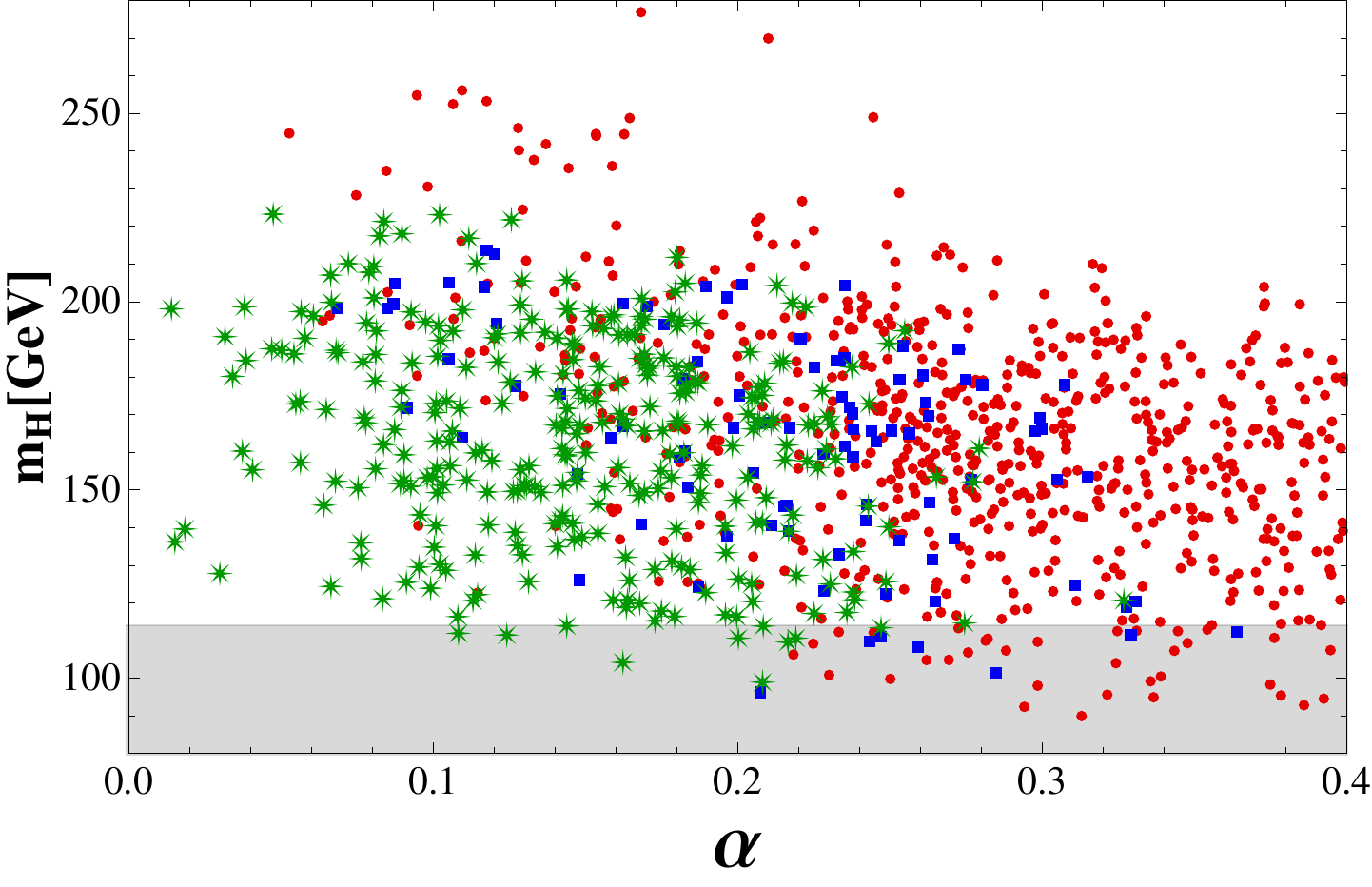}
\end{center}
\caption{\it Scatter plot of points obtained from a scan over the parameter space of MCHM$_5$.
Small red dots represent points which don't pass EWPT at $99\%$C.L., square blue dots represent
points which pass EWPT at $99\%$C.L. but not at  $90\%$ C.L., and star shape green dots represent
points which pass EWPT at $90\%$C.L.. The region below the LEP bound ($m_H < 114\ {\rm GeV}$) is shaded.}
\label{fig:mchm5mhalpha}
\end{figure} 

The SM quarks are embedded in bulk fermions transforming in the fundamental representation of $SO(5)$.
For each quark generation, 4 bulk fermions $\xi_{q_1}$, $\xi_{q_2}$, $\xi_u$ and $\xi_d$
in the $\mathbf{5}$ are introduced. The holographic Lagrangian for the third quark generation can be written to all orders in $s_\alpha$ and has the simple form
\bea
{\cal L}_H& = &\bar q_L \frac{\pslash}{p}\bigg[ \Pi_0^q  +s_\alpha^2
\Big(\Pi_1^{q_u} \frac{H^c (H^c)^\dagger}{H^\dagger H}+ \Pi_1^{q_d} \frac{H H^\dagger}{H^\dagger H} \Big)\bigg] q_L
+ \sum_{a=u,d} \bar a_R  \frac{\pslash}{p} \Big(\Pi_0^a +s_\alpha^2 \Pi_1^a \Big) a_R\nn \\
& + & \frac{s_{2\alpha}}{2h}(\Pi^u_M \bar q_L H^c u_R +\Pi^d_M \bar q_L H d_R+h.c.) \,.
\label{Lholomchm5}
\eea
The expression of the form factors appearing in eq.(\ref{Lholomchm5}) is reported in eq.(C.3) of \cite{Serone:2009kf}.
The top and bottom quark masses are approximately given by 
\be
\frac{M^2_{t}}{M^2_W}  \simeq  \frac{\theta |\tilde m_u-\tilde M_u^{-1}|^2
e^{2 L (m_u-m_1)}}{N_L N_{uR}}, \ \ \ \ \  
\frac{M^2_{b}}{M^2_W} \simeq  \frac{\theta |\tilde m_d-\tilde M_d^{-1}|^2
e^{2L(m_d-m_2)}}{N_L N_{dR}},
\label{masstopbot}
\ee
where
 \begin{figure}[t]
\begin{center}
\includegraphics[width=0.66\textwidth]{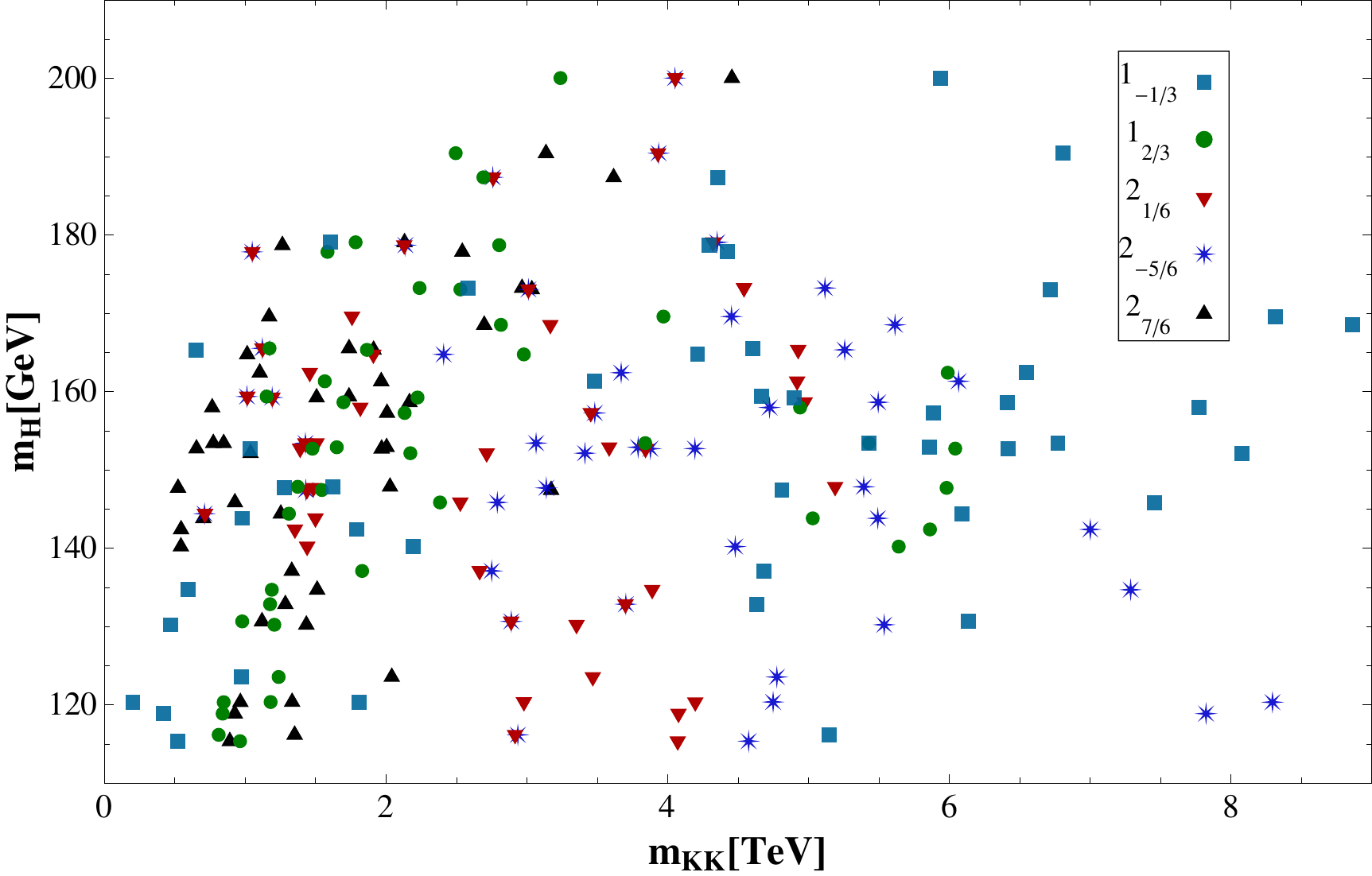}
\end{center}
\caption{\it Higgs mass $m_H$ versus the mass of the first KK resonances (before EWSB) for the points of the MCHM$_{5}$ model with $m_H>114\ {\rm GeV}$ and  $\alpha\in[0.26,0.34]$ .}
\label{fig:mchm5mkk}
\end{figure} 
 \bea
N_L & =  & \lim_{p\rightarrow 0} \frac{\Pi_0^q }{pL} = \frac 1L\sum_{i=u,d,q_1,q_2}  \int_0^L \!dy f_{iL}^2(y)\,,\nn \\
N_{uR} & = &  \lim_{p\rightarrow 0} \frac{\Pi_0^u }{pL}  = \frac 1L \int_0^L \!dy \Big(f_{uR}^2(y)+f_{q_1R}^2(y)\Big)\,, \nn \\
N_{dR} & = &  \lim_{p\rightarrow 0} \frac{\Pi_0^d }{pL}  =  \frac 1L\int_0^L \!dy \Big(f_{dR}^2(y)+f_{q_2R}^2(y)\Big)\,,
\eea
with $f_{iL,iR}(y)$ the ``holographic" wave functions of the LH/RH top and bottom quarks before EWSB. They read
\bea
f_{q_1L} & = & e^{-m_1 y}\,, \ \ f_{q_2L} = e^{-m_2 y}\,,  \ \ f_{uL} = -\tilde m_u e^{-m_1 L+m_u(L-y)}\,, \ \ f_{dL} = -\tilde m_d e^{-m_2 L+m_d(L-y)} \,,\! \nn \\
f_{uR} & = & e^{m_u y}\,,  \ \ f_{q_1R} = \frac{1}{\tilde M_u} e^{m_u L-m_1(L-y)}\,, \ \ f_{dR}  = e^{m_d y}\,, \ \  f_{q_2R} = \frac{1}{\tilde M_d} e^{m_d L-m_2(L-y)}.
\label{HoloWaveFun}
\eea
The spectrum of fermion resonances beyond the SM, before EWSB, is given by KK towers of states in the ${\bf 2}_{7/6}$,  ${\bf 2}_{-5/6}$,  ${\bf 2}_{1/6}$,  ${\bf 1}_{2/3}$ and ${\bf 1}_{-1/3}$ of $SU(2)_L\times U(1)_Y$. 
The fermion contribution  to the one-loop Higgs effective potential in this model arises only from the KK towers of the charge $+2/3$ and $-1/3$ states, and can easily be expressed in terms of the form factors appearing in eq.(\ref{Lholomchm5}). We have $V_f=V_t+V_b$, with
\be
V_i  = -2 N_c \int  \!\!\! \frac{d^4p}{(2\pi)^4}  \log\bigg[\Big(1+s_{\alpha}^2\frac{ \Pi_1^{q_i}}{\Pi_0^q}\Big)\Big(1+s_{\alpha}^2\frac{ \Pi_1^{i}}{\Pi_0^i}\Big)-s_{2\alpha}^2\frac{(\Pi^i_M)^2}{8\Pi_0^q\Pi_0^i}\bigg] \,, \ \ \ \ \ i=t,b \,.\label{VfUp}
\ee
The total Higgs potential is 
\be
V_{tot} = V_g + V_t+V_b\,,
\ee
with $V_g$ given in eq.(\ref{Potso5gauge}).

 \begin{figure}
\begin{center}
\includegraphics[width=0.6\textwidth]{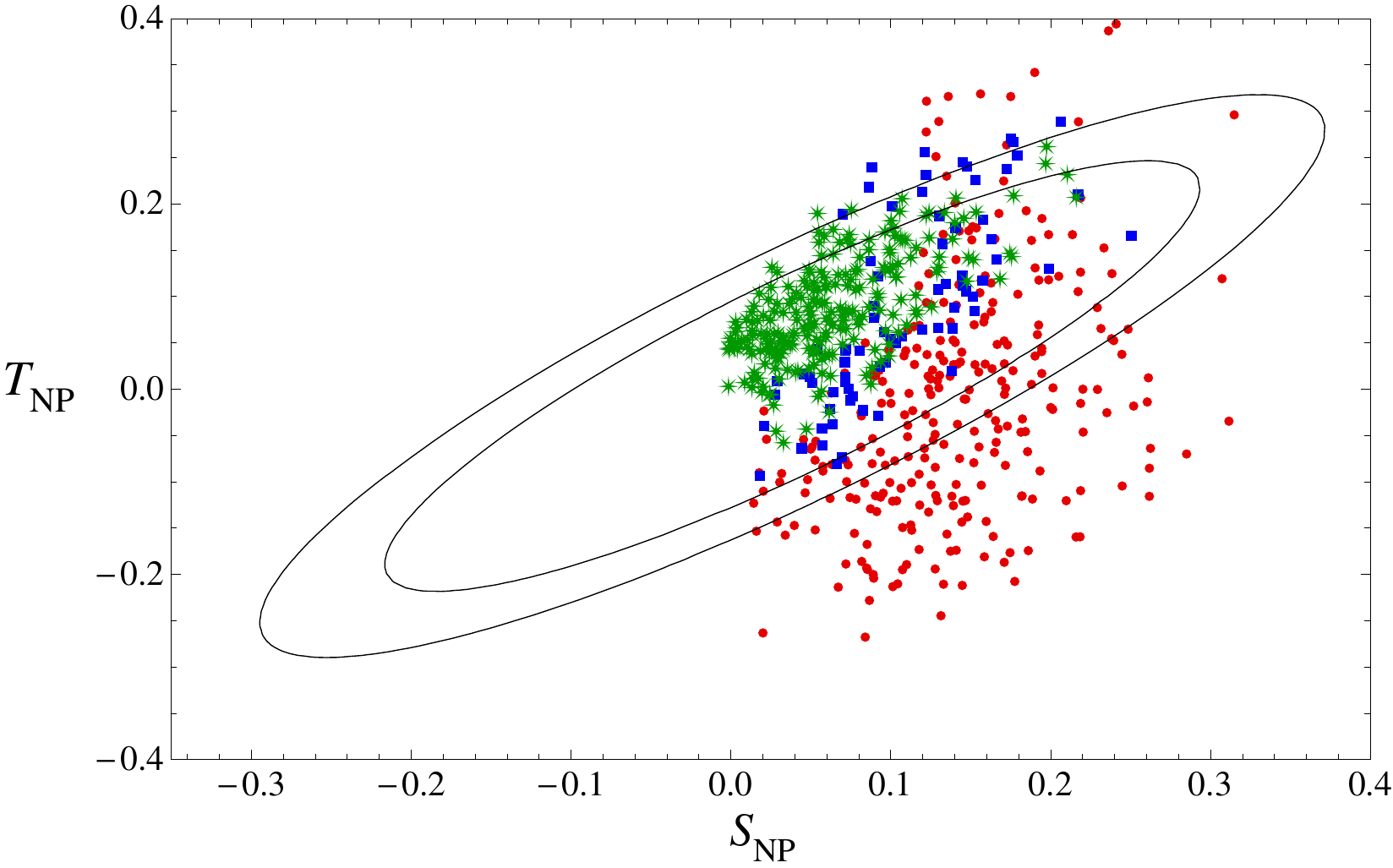}
\end{center}
\caption{\it Scatter plot of points in the MCHM$_5$ model with  $m_H> 114\ {\rm GeV}$ and  projected on the $T_{NP}$-$S_{NP}$ plane. We have set $M_{H,eff} = 120\ {\rm GeV}$. Star shape red dots represent points which don't pass EWPT at $99\%$ C.L., small blue dots represent  points which pass EWPT at $99\%$C.L. but not at $90\%$ C.L., and big green dots represent  points which pass EWPT at $90\%$ C.L.. The big and small ellipses correspond to $99\%$ and $90\%$ C.L. respectively.}
\label{fig:mchm5ST}
\end{figure}

The tree-level contribution to $\delta g_{b}$ at leading order in an expansion in $\alpha$ is
\be
\delta g_{b}  =  \frac{\alpha^2}{2 N_{L}}\sum_{i=u,d,q_1,q_2} \left(T_{3,i}^R - T_3^L\right)\int_0^L dy \Big[\frac{y}{L} ( f_{iL}^2+f_{iL} \delta f_{iL}) + \frac 12 \delta f_{iL}^2 \Big]\,,
\label{deltagb}
\ee
where $T_3^L=-1/2$, $T_{3,i}^R$ is the $SU(2)_R$ isospin of the corresponding bidoublet component where the $b_L$ lives, $f_{iL}$ are the holographic wave functions of the bidoublet components of the 5D multiplets, reported in the first line of eq.(\ref{HoloWaveFun}), and $\delta f_{iL} = f_{iL}-f_{iL}^s$, with $f_{iL}^s$ the holographic wave functions of the singlet components of the 5D multiplets. 
Only multiplets where $T_3^R\neq T_3^L$ contribute to $\delta g_{b}$, as expected \cite{Agashe:2006at}. We have $T_{3,q_1}^R = T_{3,u}^R= -1/2$,  $T_{3,q_2}^R = T_{3,d}^R= 1/2$, so that only the latter contribute to $\delta g_{b}$. We also have $f_{q_2L}^s=f_{q_2L}$  and $f_{dL}^s = \tilde M_d/\tilde m_d f_{dL}$.

\subsection{Results}

The results of our numerical scan are summarized in figs.\ref{fig:mchm5mhalpha}, \ref{fig:mchm5mkk} and \ref{fig:mchm5ST}. 
The randomly chosen input parameters are $m_u$, $m_d$, $m_1$, $m_2$, $\tilde m_u$, $\tilde m_d$, $\theta$ and $\theta^\prime$.
The remaining two parameters $\tilde M_u$ and $\tilde M_d$ are fixed by the top and bottom mass formulas.
Demanding a small $\delta g_b$ at tree-level requires $m_2 L \gtrsim 1$,
as can be verified by using eq.~(\ref{deltagb}). We have scanned the parameter space
over the region  $m_u L\in[-3,3]$, $m_d L\in[-5,2.5]$, $m_1 L\in[-2,2]$, $m_2 L\in[2.2,4.5]$,
$\tilde m_u\in[-2.3,4.1]$, $\tilde m_d\in[-3.5,4]$, $\theta\in[17,27]$,
$\theta^\prime\in[14,26]$. As can be seen in fig.\ref{fig:mchm5mhalpha}, the EWPT
constraints are now milder, with $\alpha\simeq 1/3$. The Higgs is still light,
but now masses up to 200 GeV can be reached in the less-tuned region $\alpha \simeq 1/3$.
There is no definite pattern for the lightest exotic particles. 
As far as collider physics is concerned, this model is the most interesting one,
having a scale of new physics lower than that associated to the FBKT models and fermion states
below the TeV scale.  As expected, $b_R$ is always mostly elementary, while $q_L$ and $t_R$ typically show a sizable degree of compositeness.
Depending on the region in parameter space, $q_L$ can be semi-composite and $t_R$ mostly composite or the other way around, with
$q_L$ mostly composite and $t_R$ semi-composite.

\section{Comments on the EWPT}

We give in this section a rough qualitative picture of how the EWPT are passed in
each model. We do not try to perform here an analysis of how EWSB selects the fermion
mass spectra reported in figs.\ref{fig:bkt10mkk}, \ref{fig:bkt5mkk} and \ref{fig:mchm5mkk},
but rather we take these spectra for granted. 
An obvious feature common to all the models is $S_{NP}>0$, coming from the dominant
tree-level contribution (\ref{Streeso5}). As well-known, given $S_{NP}>0$, the EWPT favours
a light Higgs and models where $T_{NP}>0$ rather than $T_{NP}<0$. As far as $\delta g_b$
is concerned, by comparing eq.~(\ref{epsTheo}) with eq.~(\ref{epsExp}), one finds that
models where $\delta g_{b,NP}<0$ are slightly favoured with respect to the ones with
$\delta g_{b,NP}>0$.
Let us now turn to each model separately.

In the FBKT$_{10}$ model, the lightest resonances before EWSB are the first KK states of
the exotic ${\bf 1}_{5/3}$ and ${\bf 2}_{7/6}$ towers, see fig.\ref{fig:bkt10mkk}.
The charge $5/3$ states do not contribute to $\delta g_{b,NP}$ at one-loop level, and
their contribution to $T_{NP}$ is negligible with respect to the one given by the charge 2/3 states (compare eq.(\ref{TBD7}) with eq.(\ref{T53})).
The latter contribution to $T_{NP}$ is always negative. The FBKT$_{10}$ model features an $SU(2)_L$  triplet state, whose tree-level
contribution to $\delta g_{b,NP}$ can be sizable and can play an important role in the EWPT. The
total combination of these effects does not allow to have large enough values of $\alpha$, which
is then constrained to be at most $\alpha\sim 0.2$.

In the FBKT$_5$ model, the lightest resonances before EWSB are the first KK states of the
${\bf 1}_{-1/3}$ tower. As we already remarked, such states have a small overlap with the bottom
quark and hence a negligible contribution to $S_{NP}$, $T_{NP}$ and $\delta g_{b,NP}$.  The next
to lightest resonances are the two towers of states in the ${\bf 2}_{1/6}$ and ${\bf 2}_{7/6}$,
which have comparable Yukawa couplings with the top quark.
By $SO(4)$ symmetry, the net contributions to $T_{NP}$ and $\delta g_{b,NP}$ of these two
states tend to compensate each other. Since typically the ${\bf 2}_{7/6}$ states are lighter
than the ${\bf 2}_{1/6}$ ones, we get  a net negative contribution to $T_{NP}$ and a negligible
contribution to $\delta g_{b,NP}$, since the doublets contribution to $\delta g_{b,NP}$ is
suppressed (see eqs.(\ref{gbBD1}) and (\ref{gbBD7})). In this situation, $\delta g_b$ is
sub-dominant and we get the quite standard behaviour depicted in fig.\ref{fig:bkt5ST}, with
$S_{NP}>0$ and $T_{NP}<0$.

The MCHM$_5$ model is the most interesting and complicated to analyze, given also the more intricate pattern of fermion spectrum depicted in fig.\ref{fig:mchm5mkk}.
The tree-level contribution to $\delta g_b$ is typically positive but small. The novelty of this model with respect to the FBKT ones is the appearance of configurations satisfying EWPT with a sizable positive $T_{NP}$, see fig.\ref{fig:mchm5ST}. This is related to two features appearing in the MCHM$_5$ model. The first is the possibility of having  a moderate hierarchy between the Yukawa couplings $\lambda_1$ and $\lambda_7$ of the lightest states  ${\bf 2}_{1/6}$ and ${\bf 2}_{7/6}$ with the top quark. It arises thanks to the presence of four bulk fields and IR mass terms, that lead to a larger $SO(5)$ symmetry breaking with respect to the FBKT models. As in the FBKT$_5$ model,  the ${\bf 2}_{7/6}$ states are lighter than the ${\bf 2}_{1/6}$ ones (with the singlet ${\bf 1}_{1/3}$ considerably heavier than both), but  it often happens that $\lambda_1 > \lambda_7$ 
resulting in a net dominance of the ${\bf 2}_{1/6}$ state with respect to the ${\bf 2}_{7/6}$ (a similar pattern arises in a specific region in parameter space of the warped model studied in \cite{Panico:2008bx}).
The total result is a sizable $T_{NP}>0$ and a negligibly small $\delta g_{b,NP}$. 
The second feature is the possibility of having light singlet ${\bf 1}_{2/3}$ states (compare figs.\ref{fig:bkt10mkk} and \ref{fig:bkt5mkk} with fig.\ref{fig:mchm5mkk}). In the FBKT models, the ${\bf 1}_{2/3}$ states always have $(++)$ or $(--)$ b.c. and 
are hence heavy, while in the MCHM$_5$ model, due to the presence of more bulk fields and IR mass terms, they have mixed b.c. and can be light.
Their presence is important, because they positively contribute to $T_{NP}$. Taken alone, the singlets would also give rise to unacceptably positive and large contributions to $\delta g_{b,NP}$, but
it turns out that the net effect of the ${\bf 2}_{1/6}$, ${\bf 2}_{7/6}$ and ${\bf 1}_{2/3}$ states is to keep $\delta g_{b,NP}$ small, while having $T_{NP}$ positive and sizable.
We do not exclude that other patterns may exist, where the same desired configuration of $T_{NP}>0$ and $\delta g_{b,NP} \simeq 0$ is obtained.

\section{Conclusions}

We have constructed three different composite Higgs/GHU models in flat space with large BKT, based on the minimal custodially-symmetric $SO(5)\times U(1)_X$ gauge group, and we have shown that EWSB and EWPT are compatible in these models.  We stress that model building in this context is significantly simpler than in warped space.

The Higgs is predicted to be light with a mass $m_H \leq 200$ GeV.
The lightest new-physics particles are colored fermions with a mass as low as about $500$ GeV in the MCHM$_5$ model and 1 TeV in the FBKT models.
Their electroweak quantum numbers depend on the model and on the region in parameter space, but they are always particles with electric charges -1/3, +2/3 or 
+5/3.

The next step in constructing fully realistic models would be the addition of the light two quark generations, leptons, and flavour in general. 
We expect that the typical known patterns of flavour physics in warped space, such as the so-called RS-GIM, should also be captured by our effective flat space description.
Indeed, in presence of large BKT, the cut-off of the theory becomes effectively a function of the position in the internal space and is maximal
at the UV brane, with the SM fields becoming more elementary (peaked at the UV brane at $y=0$) and the KK states more composite (peaked at the IR brane at $y=L$). In this way, otherwise too large flavour-changing operators might be naturally suppressed.
It would be very interesting to study this issue in detail and see whether and to what extent this expectation is valid.

The very broad collider signatures of our models completely fall into those of composite Higgs/warped GHU models.
The correct EWSB pattern in all composite Higgs/GHU models constructed so far  (warped or flat,
with $SO(5)$ or $SU(3)$ gauge groups) seems to indicate that the lightest (below TeV)
new physics states beyond the SM should be fermionic colored particles, with model-dependent
$SU(2)_L\times U(1)_Y$ quantum numbers. Of course, this generic prediction cannot be seen as a
``signature" of composite Higgs/GHU models.
More specific predictions are the expected sizable deviations to the SM Higgs-gauge
couplings or to the SM top couplings, but at this stage of the LHC run these are details that cannot be detected in the short term.

\section*{Acknowledgments}

G.P. would like to thank Andrea Wulzer for useful discussions and comments. M.S. thanks Eduardo Pont\'on and Jos\'e Santiago for a useful correspondence and especially Alessandro Strumia for having provided us with the updated numerical coefficients entering in our fit.

\appendix

\section{One-loop fermion contribution to the $S$, $T$ parameters and the $Z b_L \ov b_L$ vertex}

\label{app:STbbloop}

We collect in this appendix the one-loop fermion contribution to $T$, $S$ and $\delta g_b$
in particular limits where relatively simple analytic expressions are available. 
This is motivated by the fact that often in our models one or two fermion states are significantly lighter than
the others and dominate the loop corrections. The SM quantum numbers of these light fermion states vary along the parameter space and thus it can be useful to list the single fermion contribution to $T$, $S$ and $\delta g_b$.  We compute the one-loop contribution to $\delta g_b$ in the approximation in which the external momentum of the $Z$ is set to zero (see \cite{Anastasiou} for a more general computation).
The contribution of the first two light quark generations, including their KK towers, given their light masses and small 
Yukawa couplings with the KK modes, is expected to be negligible. 
We have actually checked that even the fermion mixing in the bottom sector is negligible, so that only the charge +2/3 states mixing with the top quark should be considered. 
We define in what follows by  $T_{NP}$, $S_{NP}$ and $\delta g_{b,NP}$ the fermion one-loop contribution given by new physics only, with the SM contribution subtracted:
\be
T_{NP}= T-T_{SM}\,, \ \ \ \  S_{NP}= S-S_{SM}, \ \ \ \ \ \delta g_{b,NP}= \delta g_b-\delta g_{b,SM}\,,
\label{STgbDef}
\ee
where
\bea
g_{b,SM} & = & -\frac 12 +\frac 13 s_W^2 \,, \ \ \ \  T_{SM}   \simeq   \frac{N_c r}{16\pi s^2_W} \,, \ \ \ \ \  
S_{SM} =  \frac{N_c}{18\pi} \left(3+\log \left(\frac{M_b^2}{M_t^2}\right)\right)\,, \nn \\
\delta g_{b,SM} & = & \frac{\alpha_{em}}{16\pi s^2_W} \frac{r(r^2-7r+6+(2+3r)\log r)}{(r-1)^2}\,, \ \ \ r\equiv \frac{M_t^2}{M_W^2}\,,
\eea
$s_W\equiv \sin\theta_W$, and $M_t$ is the pole top mass, $M_t = 173.1$ GeV \cite{:2009ec}.

We do not exploit the full $SO(5)$ symmetry underlying our model
and classify the new fermion states by their SM quantum numbers. In this way, the explicit $SO(5)$ symmetry breaking effects
due to the UV b.c., that can be sizable, are taken into account and more reliable expressions are obtained.
For simplicity, we take in the following all Yukawa couplings to be real, the extension to complex ones being straightforward.

\subsection{Singlet with $Y=2/3$}

The simplest situation arises when the top quark mixes with just one SM singlet vector-like fermion $X$ with hypercharge $Y=2/3$.
The two possible Yukawa couplings are
\be
{\cal L} \supset y_t \bar q_L H^c t_R + y_X \bar q_L H^c X_R + h.c. \rightarrow \lambda_t \bar t_L t_R + \lambda_X \bar t_L X_R + h.c.\,,
\ee
where here and in the following we use the notation that $\lambda_i = y_i v/\sqrt{2}$ is the mass parameter corresponding to the Yukawa coupling $y_i$. The $\lambda_i$  are assumed to be small with respect to the vector-like mass $M_X$ of the new exotic fermions. By using standard techniques and keeping the leading order terms in the $\lambda_i/M_X$ expansion, we get
\bea
T_{NP} & =  & \frac{N_c \lambda_X^2 \Big(2\lambda_t^2\log\big(\frac{M_X^2}{\lambda^2_t}\big)+\lambda_X^2-2\lambda_t^2\Big)}{16\pi  s^2_W M_W^2 M_X^2} \,, \label{TSinglet} \\
S_{NP} & = & \frac{N_c \lambda_X^2 \Big(2\log\big(\frac{M_X^2}{\lambda^2_t}\big)-5\Big)}{18\pi M_X^2} \,,  \label{SSinglet} \\
\delta g_{b,NP} & = & \frac{\alpha_{em} \lambda_X^2 \Big(2\lambda_t^2 \log\big(\frac{M_X^2}{\lambda^2_t}\big)+\lambda_X^2-2\lambda_t^2\Big)}{ 16 \pi s_W^2 M_W^2 M_X^2}\,, \label{gbSinglet}
\eea
in agreement with \cite{Carena:2006bn,Barbieri:2007bh}.
For simplicity,  in eq.(\ref{gbSinglet}) we have only reported the leading order
terms in the limit $\lambda_i/M_W \gg 1$. The top mass is given by
\be
M_t \simeq \lambda_t  \Big(1- \frac{\lambda_X^2}{2M_X^2}\Big)\,.
\ee 
As can be seen from eqs.(\ref{TSinglet})-(\ref{gbSinglet}), for a sufficiently large $M_X$, $T_{NP}$ and $\delta g_{b,NP}$ are closely related and positive (like $S_{NP}$).

\subsection{Doublet with $Y=1/6$}

The two possible Yukawa couplings mixing the top with a new doublet $Q_1$ with hypercharge $Y=1/6$ are
\be
{\cal L} \supset y_t \bar q_L H^c t_R + y_1 \bar Q_{1L} H^c t_R + h.c.  \rightarrow \lambda_t \bar t_L t_R + \lambda_1 \bar Q_{1uL} t_R + h.c.\,.
\ee
We find
\bea
T_{NP} & =  & \frac{N_c \lambda_1^2 \Big(6\lambda_t^2\log\big(\frac{M_1^2}{\lambda^2_t}\big)+2\lambda_1^2-9\lambda_t^2\Big)}{24\pi  s^2_W M_W^2 M_1^2} \,, \label{TBD1} \\
S_{NP} & = & \frac{N_c \lambda_1^2 \Big(4\log\big(\frac{M_1^2}{\lambda^2_t}\big)-7\Big)}{18\pi M_1^2} \,,  \label{SBD1} \\
\delta g_{b,NP} & = & \frac{\alpha_{em} \lambda_1^2 \lambda_t^2 \log\big(\frac{M_1^2}{\lambda^2_t}\big)}{ 32 \pi s_W^2 M_W^2 M_1^2}\,, \label{gbBD1}
\eea
in agreement with \cite{Carena:2006bn}.
The top mass is given by
\be
M_t \simeq \lambda_t  \Big(1- \frac{\lambda_1^2}{2M_1^2}\Big)\,.
\ee 
As can be seen from eqs.(\ref{TBD1})-(\ref{gbBD1}), for a sufficiently large $M_1$,  $T_{NP}$, $\delta g_{b,NP}$ and $S_{NP}$ are all positive. 

\subsection{Doublet with $Y=7/6$}

The two possible Yukawa couplings mixing the top with a new doublet $Q_7$ with hypercharge $Y=7/6$ are
\be
{\cal L} \supset y_t \bar q_L H^c t_R + y_7 \bar Q_{7L} H t_R + h.c.  \rightarrow \lambda_t \bar t_L t_R + \lambda_7 \bar Q_{7dL} t_R + h.c.\,.
\ee
We find
\bea
T_{NP} & =  & -\frac{N_c \lambda_7^2 \Big(6\lambda_t^2\log\big(\frac{M_7^2}{\lambda^2_t}\big)-2\lambda_7^2-9\lambda_t^2\Big)}{24\pi  s^2_W M_W^2 M_7^2} \,, \label{TBD7} \\
S_{NP} & = & -\frac{N_c \lambda_7^2 \Big(4\log\big(\frac{M_7^2}{\lambda^2_t}\big)-15\Big)}{18\pi M_7^2} \,,  \label{SBD7} \\
\delta g_{b,NP} & = & -\frac{\alpha_{em} \lambda_7^2 \lambda_t^2 \log\big(\frac{M_7^2}{\lambda^2_t}\big)}{ 32 \pi s_W^2 M_W^2 M_7^2}\,, \label{gbBD7}
\eea
in agreement with \cite{Carena:2006bn}.
The top mass is given by
\be
M_t \simeq \lambda_t  \Big(1- \frac{\lambda_7^2}{2M_7^2}\Big)\,.
\ee 
As can be seen from eqs.(\ref{TBD7})-(\ref{gbBD7}), for a sufficiently large $M_7$,  $T_{NP}$, $\delta g_{b,NP}$
and $S_{NP}$ are all negative.

The contributions to $T$, $S$ and $\delta g_b$ of the doublets with $Y=1/6$ and $Y=7/6$ are almost the same in magnitude, but opposite in sign.
When present together, then, there tends to be a partial cancellation among these two contributions.
In the $SO(4)$ invariant limit in which $M_1=M_7$ and $\lambda_1=\lambda_7$, their contributions to $T$ and $\delta g_b$ precisely cancel.

\subsection{Triplet with $Y=2/3$}

The two possible Yukawa couplings mixing the top with a new triplet $T$ with hypercharge $Y=2/3$ are
\be
{\cal L} \supset y_t \bar q_L H^c t_R + \sqrt{2} y_T \bar q_{L} T_R H^c  + h.c.  \rightarrow \lambda_t \bar t_L t_R + \lambda_T \bar t_{L} T_{0R} + h.c.\,,
\label{YukTriplet}
\ee
where $T_{0,R}$ is the triplet component with $T_{3L}=0$.
We find
\bea
T_{NP} & =  & \frac{N_c \lambda_T^2 \Big(18\lambda_t^2\log\big(\frac{M_T^2}{\lambda^2_t}\big)+19\lambda_T^2-30\lambda_t^2\Big)}{48\pi  s^2_W M_W^2 M_T^2} \,, \label{TTri} \\
S_{NP} & = & -\frac{N_c \lambda_T^2 \Big(4\log\big(\frac{M_T^3 \lambda_t}{\lambda^4_b}\big)-29\Big)}{18\pi M_T^2} \,,  \label{STri} \\
\delta g_{b,NP} & = & -\frac{\alpha_{em} \lambda_T^2 \Big(2\lambda_t^2 \log\big(\frac{M_T^2}{\lambda^2_t}\big)-\lambda_T^2\Big)}{ 16 \pi s_W^2 M_W^2 M_T^2}\,. \label{gbTri}
\eea
The top mass is given by
\be
M_t \simeq \lambda_t  \Big(1- \frac{\lambda_T^2}{2M_T^2}\Big)\,.
\ee 
As can be seen from eqs.(\ref{TTri})-(\ref{gbTri}),  $T_{NP}>0$ and $\delta g_{b,NP}<0$.
Contrary to the previous cases,  the bottom quark mixing cannot consistently be neglected, since the same Yukawa coupling in eq.(\ref{YukTriplet}) mixing the top with the $T_{3L}=0$ triplet component gives also a mixing between the bottom and the $T_{3L}=-1$ triplet component. This mixing is at the origin of the log term involving the bottom Yukawa coupling $\lambda_b$ in eq.(\ref{STri}), which enhances the fermion one-loop contribution to $S$ with respect to the previous cases and gives $S_{NP}<0$. 

\subsection{Doublet with $Y=7/6$ mixing with singlet with $Y=5/3$}

The two Yukawa couplings mixing a vector-like singlet $X$ with hypercharge $Y=5/3$ with a vector-like doublet $Q_7$ with $Y=7/6$ are
\be
{\cal L} \supset y_{XL} \bar Q_{7R} H^c X_L + y_{XR} \bar Q_{7L} H^c X_R + h.c.  \rightarrow \lambda_{XL} \bar Q_{7uR} X_L + \lambda_{XR} \bar Q_{7uL} X_R + h.c.\,.
\ee
In the limit in which $M_7 = M_X$, we have
\bea
T_{NP} & =  & \frac{N_c \Big(13\lambda_{XL}^4+2\lambda_{XL}^3\lambda_{XR}+18\lambda_{XL}^2\lambda_{XR}^2 +2\lambda_{XL}\lambda_{XR}^3+13\lambda_{XR}^4\Big)}{480\pi  s^2_W M_W^2 M_X^2} \,, \label{T53} \\
S_{NP} & = & \frac{N_c \Big(12\lambda_{XL}^2+79\lambda_{XL}\lambda_{XR}+12\lambda_{XR}^2\Big)}{90\pi M_X^2} \,.  \label{S53} 
\eea
Of course, $\delta g_b$ vanishes, since there is no coupling between the bottom and these states. Being given by vector-like states, eqs.(\ref{T53}) and (\ref{S53}) do not contain ``large" log's of the form $\log M/\lambda_t$.  Assuming equality of masses and Yukawa's,  the contribution to $T$ in eq.(\ref{T53}) is suppressed with respect to the other contributions previously determined.


\thebibliography{99}

 \bibitem{early}
D.~B.~Fairlie,
Phys.\ Lett.\ B {\bf 82} (1979) 97;
J.\ Phys.\ G {\bf 5} (1979) L55;\\
N.~S.~Manton,
Nucl.\ Phys.\ B {\bf 158} (1979) 141;\\
P.~Forgacs, N.~S.~Manton,
Commun.\ Math.\ Phys.\  {\bf 72} (1980) 15; \\
Y. Hosotani,
Phys.\ Lett.\ B {\bf 126} (1983) 309;
ibid. {\bf 129} (1983) 193;
Ann.\ Phys.\ {\bf 190} (1989) 233.

 \bibitem{Randall:1999ee}
  L.~Randall and R.~Sundrum,
  Phys.\ Rev.\ Lett.\  {\bf 83} (1999) 3370
  [arXiv:hep-ph/9905221];
  Phys.\ Rev.\ Lett.\  {\bf 83} (1999) 4690
  [arXiv:hep-th/9906064].

  \bibitem{ArkaniHamed:2000ds}
  N.~Arkani-Hamed, M.~Porrati and L.~Randall,
  JHEP {\bf 0108} (2001) 017
  [arXiv:hep-th/0012148]; \\
  R.~Rattazzi and A.~Zaffaroni,
  JHEP {\bf 0104} (2001) 021
  [arXiv:hep-th/0012248]; \\
   M.~Perez-Victoria,
  JHEP {\bf 0105} (2001) 064
  [arXiv:hep-th/0105048].
  
  \bibitem{Contino:2003ve}
  R.~Contino, Y.~Nomura and A.~Pomarol,
  Nucl.\ Phys.\  B {\bf 671} (2003) 148
  [arXiv:hep-ph/0306259].
  
\bibitem{Agashe:2004rs}
  K.~Agashe, R.~Contino and A.~Pomarol,
  Nucl.\ Phys.\  B {\bf 719} (2005) 165
  [arXiv:hep-ph/0412089].

 \bibitem{Ads-cft}
  J.~M.~Maldacena,
  Adv.\ Theor.\ Math.\ Phys.\  {\bf 2} (1998) 231
  [Int.\ J.\ Theor.\ Phys.\  {\bf 38} (1999) 1113]
  [arXiv:hep-th/9711200];  \\ 
   S.~S.~Gubser, I.~R.~Klebanov and A.~M.~Polyakov,
  Phys.\ Lett.\ B {\bf 428} (1998) 105
  [arXiv:hep-th/9802109]; \\ 
   E.~Witten,
  Adv.\ Theor.\ Math.\ Phys.\  {\bf 2} (1998) 253
  [arXiv:hep-th/9802150].

 \bibitem{Kaplan}
  D.~B.~Kaplan and H.~Georgi,
  Phys.\ Lett.\  B {\bf 136} (1984) 183; 
  M.~J.~Dugan, H.~Georgi, D.~B.~Kaplan,
  Nucl.\ Phys.\  {\bf B254 } (1985)  299.

  \bibitem{Giudice:2007fh}
  G.~F.~Giudice, C.~Grojean, A.~Pomarol and R.~Rattazzi,
  JHEP {\bf 0706} (2007) 045
  [arXiv:hep-ph/0703164].

 \bibitem{Contino:2006nn}
  R.~Contino, T.~Kramer, M.~Son and R.~Sundrum,
  JHEP {\bf 0705} (2007) 074
  [arXiv:hep-ph/0612180].

 \bibitem{Luty:2003vm}
  M.~A.~Luty, M.~Porrati and R.~Rattazzi,
  JHEP {\bf 0309} (2003) 029
  [arXiv:hep-th/0303116].

\bibitem{Barbieri:2003pr}
  R.~Barbieri, A.~Pomarol and R.~Rattazzi,
  Phys.\ Lett.\ B {\bf 591} (2004) 141
  [arXiv:hep-ph/0310285].

  \bibitem{Contino:2004vy}
  R.~Contino and A.~Pomarol,
  JHEP {\bf 0411} (2004) 058
  [arXiv:hep-th/0406257].

 \bibitem{Panico:2007qd}
  G.~Panico and A.~Wulzer,
  JHEP {\bf 0705} (2007) 060
  [arXiv:hep-th/0703287].

      \bibitem{Peskin:1991sw}
  M.~E.~Peskin and T.~Takeuchi,
  Phys.\ Rev.\ D {\bf 46} (1992) 381; \\
  R.~Barbieri, A.~Pomarol, R.~Rattazzi and A.~Strumia,
  Nucl.\ Phys.\ B {\bf 703}, 127 (2004)
  [arXiv:hep-ph/0405040].
  
     \bibitem{delAguila:2008iz}
  F.~del Aguila {\it et al.},
  Eur.\ Phys.\ J.\  C {\bf 57} (2008) 183
  [arXiv:0801.1800 [hep-ph]].
  
  \bibitem{Agashe:2003zs}
  K.~Agashe, A.~Delgado, M.~J.~May and R.~Sundrum,
  JHEP {\bf 0308} (2003) 050
  [arXiv:hep-ph/0308036].
  
  \bibitem{Agashe:2006at}
  K.~Agashe, R.~Contino, L.~Da Rold and A.~Pomarol,
  Phys.\ Lett.\  B {\bf 641} (2006) 62
  [arXiv:hep-ph/0605341].

 \bibitem{Carena:2006bn}
  M.~S.~Carena, E.~Ponton, J.~Santiago and C.~E.~M.~Wagner,
  Nucl.\ Phys.\  B {\bf 759} (2006) 202
  [arXiv:hep-ph/0607106]; 
  Phys.\ Rev.\  D {\bf 76} (2007) 035006
  [arXiv:hep-ph/0701055].

 \bibitem{Contino:2006qr}
  R.~Contino, L.~Da Rold and A.~Pomarol,
  Phys.\ Rev.\  D {\bf 75} (2007) 055014
  [arXiv:hep-ph/0612048].

   \bibitem{Medina:2007hz}
  A.~D.~Medina, N.~R.~Shah and C.~E.~M.~Wagner,
  Phys.\ Rev.\  D {\bf 76} (2007) 095010
  [arXiv:0706.1281 [hep-ph]].

\bibitem{Panico:2008bx}
  G.~Panico, E.~Ponton, J.~Santiago and M.~Serone,
  Phys.\ Rev.\  D {\bf 77} (2008) 115012
  [arXiv:0801.1645 [hep-ph]].

  \bibitem{Serone:2009kf}
  M.~Serone,
  New J.\ Phys.\  {\bf 12} (2010) 075013
  [arXiv:0909.5619 [hep-ph]].

   \bibitem{Panico:2005dh}
  G.~Panico, M.~Serone and A.~Wulzer,
  Nucl.\ Phys.\ B {\bf 739} (2006) 186
  [arXiv:hep-ph/0510373]; 
  Nucl.\ Phys.\  B {\bf 762} (2007) 189
  [arXiv:hep-ph/0605292].

  \bibitem{Carena:2002me}
  M.~S.~Carena, T.~M.~P.~Tait and C.~E.~M.~Wagner,
  Acta Phys.\ Polon.\  B {\bf 33} (2002) 2355
  [arXiv:hep-ph/0207056].

 \bibitem{Scrucca:2003ra}
C.~A.~Scrucca, M.~Serone, L.~Silvestrini,
Nucl.\ Phys.\ B {\bf 669} (2003) 128
[hep-ph/0304220].

\bibitem{Georgi:2000ks}
  H.~Georgi, A.~K.~Grant and G.~Hailu,
  Phys.\ Lett.\  B {\bf 506} (2001) 207
  [arXiv:hep-ph/0012379].

 \bibitem{Bamert:1996px}
  P.~Bamert, C.~P.~Burgess, J.~M.~Cline, D.~London and E.~Nardi,
  Phys.\ Rev.\  D {\bf 54} (1996) 4275
  [arXiv:hep-ph/9602438].
  
  \bibitem{Altarelli:1990zd}
  G.~Altarelli and R.~Barbieri,
  Phys.\ Lett.\  B {\bf 253} (1991) 161;  G.~Altarelli, R.~Barbieri and S.~Jadach,
  Nucl.\ Phys.\  B {\bf 369} (1992) 3
  [Erratum-ibid.\  B {\bf 376} (1992) 444]; 
  G.~Altarelli, R.~Barbieri and F.~Caravaglios,
  Nucl.\ Phys.\ B {\bf 405} (1993) 3.

\bibitem{Agashe}
  K.~Agashe and R.~Contino,
  Nucl.\ Phys.\ B {\bf 742} (2006) 59
  [arXiv:hep-ph/0510164].
  
\bibitem{Barbieri:2007bh}
  R.~Barbieri, B.~Bellazzini, V.~S.~Rychkov and A.~Varagnolo,
  Phys.\ Rev.\  D {\bf 76}, 115008 (2007)
  [arXiv:0706.0432 [hep-ph]].

\bibitem{:2004qh}
  arXiv:hep-ex/0412015,
  arXiv:hep-ex/0612034.
   
 \bibitem{Anastasiou}
  C.~Anastasiou, E.~Furlan and J.~Santiago,
  Phys.\ Rev.\  D {\bf 79} (2009) 075003
  [arXiv:0901.2117 [hep-ph]].

  \bibitem{:2009ec}
    [Tevatron Electroweak Working Group and CDF and D0 Collaboration],
  arXiv:0903.2503 [hep-ex].
\end{document}